\documentclass[epj]{svjour}

\usepackage[dvips]{graphicx}
\usepackage{epsfig}

\renewcommand{\thefootnote}{\alph{footnote}}  

\def\be{\begin{equation}}
\def\ee{\end{equation}}
\def\bea{\begin{eqnarray}}
\def\eea{\end{eqnarray}}

\def\kcnn{K^{\pm} \rightarrow \pi^\pm \pi^0 \pi^0}
\def\kccc{K^{\pm} \rightarrow \pi^\pm \pi^+ \pi^-}
\def\p0p0{\pi^0 \pi^0}
\def\pp{\pi^+ \pi^-}
\def\mmm{M_{00}}
\def\mm2{M_{00}^2}

\def\statsys{\ref{cabisi_noskip_atomfree_10}--\ref{bernfit_chpt_noskip_atomfree_10}}

\def \fitA {CI}
\def \fitB {CI_A}
\def \fitC {CI^{\chi}}
\def \fitD {CI^{\chi}_A}
\def \fitE {BB}
\def \fitF {BB_A}
\def \fitG {BB^{\chi}}
\def \fitH {BB^{\chi}_A}

\def \geve     {GeV}
\def \gevp     {GeV/$c$}
\def \gevm     {GeV/$c^2$}
\def \gevmsq {(GeV/$c^2$)$^2$}

\usepackage{array}

\usepackage{amscd,amsmath,amssymb}

\begin{document}


\title{Determination of the S-wave $\pi\pi$ scattering lengths from a study of $\kcnn$ decays}

\author{
 J.R.~Batley\inst{1} \and
 A.J.~Culling\inst{1} \and
 G.~Kalmus\inst{1} \and
 C.~Lazzeroni\inst{1}\footnotemark[1] \and
 D.J.~Munday\inst{1} \and
 M.W.~Slater\inst{1}\footnotemark[1] \and
 S.A.~Wotton 
\inst{1} \and  R.~Arcidiacono\inst{2}\footnotemark[3] \and
 G.~Bocquet\inst{2} \and
 N.~Cabibbo\inst{2}\footnotemark[4]
% ...
\inst{20}}

\institute{{\em \small Cavendish Laboratory, University of Cambridge,
Cambridge, CB3 0HE, UK$\,$\footnotemark[2]} 
\and
{\em \small CERN, CH-1211 Gen\`eve 23, Switzerland} 
\and
{\em \small Joint Institute for Nuclear Research, 141980 Dubna,
Moscow region, Russia} 
\and
{\em \small The Enrico Fermi Institute, The University of Chicago,
Chicago, IL 60126, USA}
\and
{\em \small Department of Physics and Astronomy, University of
Edinburgh, JCMB King's Buildings, Mayfield Road, Edinburgh, EH9 3JZ, UK} 
\and
{\em \small Dipartimento di Fisica dell'Universit\`a e Sezione
dell'INFN di Ferrara, I-44100 Ferrara, Italy} 
\and
{\em \small Dipartimento di Fisica dell'Universit\`a e Sezione
 dell'INFN di Firenze, I-50019 Sesto Fiorentino, Italy} 
\and
{\em \small Sezione dell'INFN di Firenze, I-50019 Sesto Fiorentino, 
Italy} 
\and
{\em \small Institut f\"ur Physik, Universit\"at Mainz, D-55099
 Mainz, Germany$\,$\footnotemark[17]} 
\and
{\em \small Department of Physics and Astronomy, Northwestern
University, Evanston, IL 60208, USA}
\and
{\em \small Dipartimento di Fisica dell'Universit\`a e Sezione
dell'INFN di Perugia, I-06100 Perugia, Italy} 
\and
{\em \small Sezione dell'INFN di Perugia, I-06100 Perugia, Italy} 
\and
{\em Sezione dell'INFN di Pisa, I-56100 Pisa, Italy} 
\and
{\em Scuola Normale Superiore e Sezione dell'INFN di Pisa, I-56100
Pisa, Italy} 
\and
{\em Dipartimento di Fisica dell'Universit\`a e Sezione dell'INFN di
Pisa, I-56100 Pisa Italy} 
\and
{\em \small DSM/IRFU - CEA Saclay, F-91191 Gif-sur-Yvette, France} 
\and
{\em \small Fachbereich Physik, Universit\"at Siegen, D-57068
 Siegen, Germany$\,$\footnotemark[23]} 
\and
{\em \small Sezione dell'INFN di Torino, I-10125 Torino, Italy} 
\and
{\em \small Dipartimento di Fisica Sperimentale dell'Universit\`a e
Sezione dell'INFN di Torino, I-10125 Torino, Italy} 
\and
{\em \small \"Osterreichische Akademie der Wissenschaften, Institut
f\"ur Hochenergiephysik, A-10560 Wien, Austria$\,$\footnotemark[26]} 
}

\date{ 
\bf\underline{Published in The European Physical Journal C: Volume 64, Issue 4 (2009), Page 589}
}

\abstract{
We report the results from a study of the full sample of
$\sim 6.031 \times 10^7$ $\kcnn$ decays recorded by the NA48/2 experiment at
the CERN SPS. As first observed in this experiment,
the $\p0p0$ invariant mass $(M_{00})$ distribution shows a cusp-like
anomaly in the region around $M_{00}= 2m_+$, where $m_+$ is the charged pion
mass. This anomaly has been interpreted as an effect due mainly to the
final state charge exchange scattering process
$\pp \rightarrow \p0p0$ in $\kccc$ decay.
Fits to the $M_{00}$ distribution using two different theoretical
formulations provide the presently most precise 
determination of $a_0 - a_2$, the difference between the $\pi\pi$ S-wave
scattering lengths in the isospin $I=0$ and $I=2$ states. Higher-order $\pi\pi$
rescattering terms, included in the two formulations, allow also an independent,
though less precise, determination of $a_2$.
} 

\maketitle

\setcounter{footnote}{0}
\footnotetext[1]{University of Birmingham, Edgbaston, Birmingham,
B15 2TT, UK}
\footnotetext[2]{Funded by the UK Particle Physics and Astronomy
Research Council}
\footnotetext[3]{Dipartimento di Fisica Sperimentale
dell'Universit\`a e Sezione dell'INFN di Torino, I-10125 Torino,
Italy}
\footnotetext[4]{Universit\`a di Roma ``La Sapienza'' e Sezione
dell'INFN di Roma, I-00185 Roma, Italy}
\footnotetext[5]{Istituto di Cosmogeofisica del CNR di Torino,
I-10133 Torino, Italy}
\footnotetext[6]{Dipartimento di Fisica dell'Universit\`a e Sezione
dell'INFN di Ferrara, I-44100 Ferrara, Italy}
\footnotetext[7]{Scuola Normale Superiore, I-56100 Pisa, Italy}
\footnotetext[8]{CERN, CH-1211 Gen\`eve 23, Switzerland}
\footnotetext[9]{Faculty of Physics, University of Sofia ``St. Kl.
Ohridski'', 5 J. Bourchier Blvd., 1164 Sofia, Bulgaria}
\footnotetext[10]{Sezione dell'INFN di Perugia, I-06100 Perugia,
Italy}
\footnotetext[11]{Northwestern University, 2145 Sheridan Road,
Evanston, IL 60208, USA}
\footnotetext[12]{Centre de Physique des Particules de Marseille,
IN2P3-CNRS, Universit\'e de la M\'editerran\'ee, Marseille, France}
\footnotetext[13]{Department of Physics and Astronomy, George Mason
University, Fairfax, VA 22030, USA}
\footnotetext[14]{Dipartimento di Fisica, Universit\`a di Modena e
Reggio Emilia, I-41100 Modena, Italy}
\footnotetext[15]{Istituto di Fisica, Universit\`a di Urbino,
I-61029 Urbino, Italy}
\footnotetext[16]{Physikalisches Institut, Universit\"at Bonn,
D-53115 Bonn, Germany}
\footnotetext[17]{Funded by the German Federal Minister for
Education and research under contract 05HK1UM1/1}
\footnotetext[18]{SLAC, Stanford University, Menlo Park, CA 94025,
USA}
\footnotetext[19]{Royal Holloway, University of London, Egham Hill,
Egham, TW20 0EX, UK}
\footnotetext[20]{UCLA, Los Angeles, CA 90024, USA}
\footnotetext[21]{Laboratori Nazionali di Frascati,
I-00044 Frascati (Rome), Italy}
\footnotetext[22]{Institut de F\'isica d'Altes Energies, UAB,
E-08193 Bellaterra (Barcelona), Spain}
\footnotetext[23]{Funded by the German Federal Minister for Research
and Technology (BMBF) under contract 056SI74}
\footnotetext[24]{University of Bern, Institute for Theoretical
Physics, Sidlerstrasse 5, CH-3012 Bern, Switzerland}
\footnotetext[25]{Centro de Investigaciones Energeticas
Medioambientales y Tecnologicas, E-28040 Madrid, Spain}
\footnotetext[26]{Funded by the Austrian Ministry for Traffic and
Research under the contract GZ 616.360/2-IV GZ 616.363/2-VIII, and
by the Fonds f\"ur Wissenschaft und Forschung FWF Nr.~P08929-PHY}

\renewcommand{\thefootnote}{\arabic{footnote}}  

\section*{Introduction}
The main purpose of the  NA48/2 experiment at the CERN SPS was to search for
direct CP violation in $K^\pm$ decay to three pions 
\cite{Batley:2006tt,Batley:2006mu,Batley:2007yfa}.
The experiment used simultaneous $K^+$ and $K^-$ beams with momenta of
$60$~\gevp ~propagating through the detector along the same beam line.
Data were collected in 2003-2004, providing large samples of fully 
reconstructed $\kccc$ and $\kcnn$ decays.

From the analysis of the data collected in 2003, we have already reported
the observation of a cusp-like anomaly in the $\p0p0$ invariant mass 
$(M_{00})$ distribution of $\kcnn$ decays in the region around
$M_{00}= 2m_+$, where $m_+$ is the charged pion mass \cite{Batley:2005ax}.  
The existence of this threshold anomaly had been first predicted in
1961 by Budini and Fonda \cite{Budini:1961}, as a result of the
charge exchange scattering process $\pp \rightarrow \p0p0$ in $\kccc$
decay. These authors had also suggested that the study of this
anomaly, once found experimentally, would allow the determination of
the cross-section for $\pp \rightarrow \p0p0$ at energies very close
to threshold. However, samples of $\kcnn$ decay events available in
those years were not sufficient to observe the effect, nor was the
$M_{00}$ resolution. As a consequence, in the absence of any
experimental verification, the article by Budini and Fonda \cite{Budini:1961}
was forgotten.

More recently, Cabibbo \cite{Cabibbo:2004gq} has proposed an interpretation
of the cusp-like anomaly along the lines proposed by Budini and Fonda
\cite{Budini:1961}, but expressing the $\kcnn$ decay amplitude in terms
of the $\pp \rightarrow \p0p0$ amplitude at threshold, $a_x$. In
the limit of exact isospin symmetry $a_x$ can be written as $(a_0 - a_2)/3$,
where $a_0$ and $a_2$ are the S-wave $\pi\pi$ scattering lengths in the 
isospin $I=0$ and $I=2$ states, respectively.  

Here we report the results from a study of the final sample of 
$\sim 6.031 \times 10^7$ $\kcnn$ decays. Best fits to two independent
theoretical formulations of rescattering effects in
$\kcnn$ and $\kccc$ decays (\cite{Cabibbo:2005ez} and
\cite {Colangelo:2006va,Bissegger:2008ff}) provide a precise determination
of  $a_0 - a_2$, and an independent, though less precise, determination 
of $a_2$.

\section{Beam and detectors}
\label{beamdet}
\begin{figure*}

\begin{center}
\setlength{\unitlength}{1mm}

\resizebox{2.00\columnwidth}{!}{%
\begin{picture}(200.,90.)                       
\includegraphics[width=200mm]{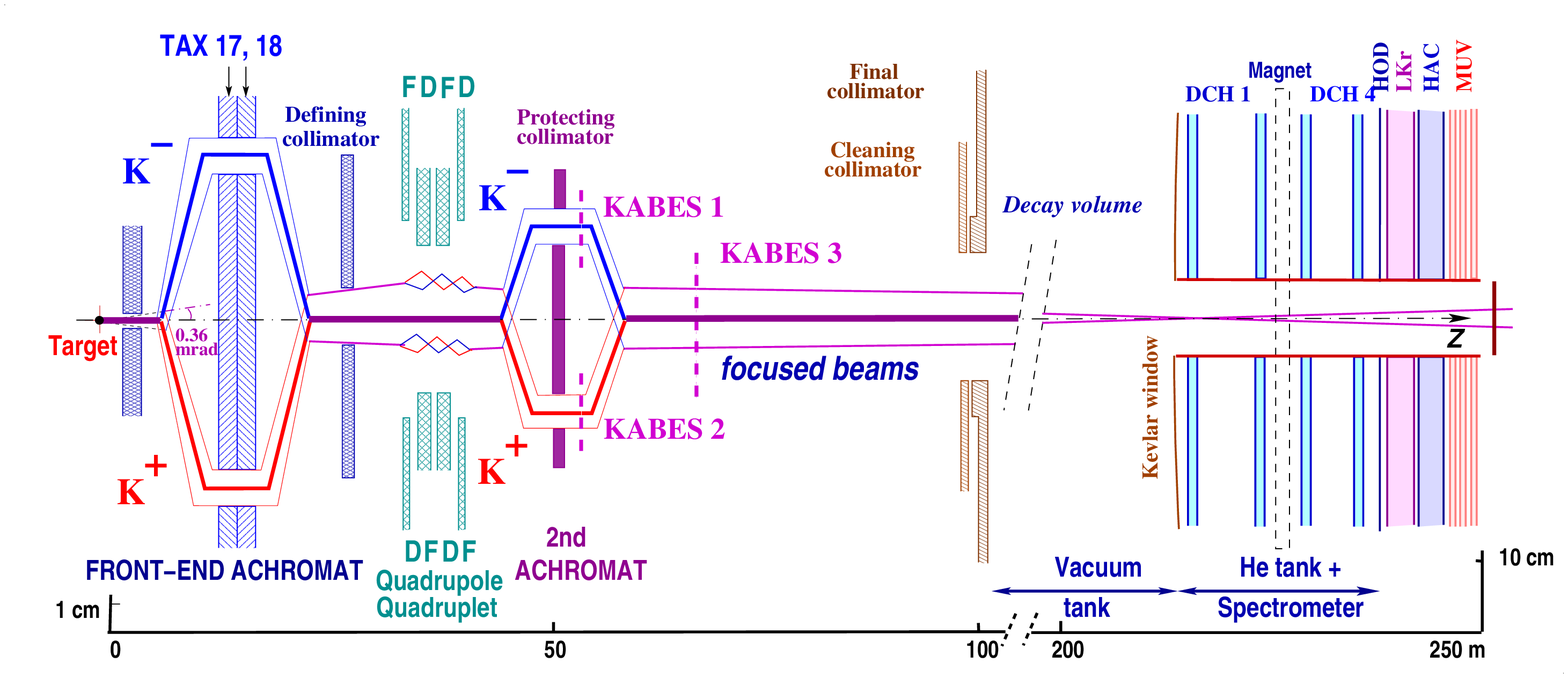}   
\end{picture}
}
\end{center}

\caption{Schematic side view of the NA48/2 beam line, decay volume and
detectors (TAX 17, 18: motorised collimators; FDFD/DFDF: focusing quadrupoles;
KABES 1-3: beam spectrometer stations (not used in this analysis); 
DCH1-4: drift chambers; HOD: scintillator hodoscope; LKr: liquid Krypton 
calorimeter; HAC: hadron calorimeter; MUV: muon veto). Thick lines indicate 
beam axes, narrow lines indicate the projections of the beam envelopes. 
Note that the vertical scales are different in the left and right part of 
the figure.}
\label{detector}

\end{figure*}

The layout of the beams and detectors is shown schematically 
in Fig. \ref{detector}.
The two simultaneous beams are produced by $400$~\gevp ~protons impinging on
a 40~cm long Be target. Particles of opposite charge with
a central momentum of $60$~\gevp ~and a momentum band of $\pm 3.8\%$ ($rms$)
produced at zero angle are selected by two systems of dipole magnets forming 
``achromats'' with null total deflection, focusing quadrupoles, muon sweepers
and collimators. With $7\times 10^{11}$ protons per pulse of $\sim 4.5$ s 
duration incident on the target the positive (negative) 
beam flux at the entrance of the decay volume
is  $3.8\times 10^{7}$ ($2.6\times 10^{7}$) particles per pulse, of which
$\sim 5.7\%$ ($\sim 4.9\%$) are $K^+$ ($K^-$). The decay volume is a 114 m 
long vacuum tank with a diameter of 1.92 m for the first 66 m, and 2.40 m 
for the rest.

A detailed description of the detector elements is available in \cite{Fanti:2007vi}.
Charged particles from $K^\pm$ decays are measured by a magnetic spectrometer 
consisting of four drift chambers (DCH1--DCH4, denoted collectively as DCH) 
and a large-aperture dipole magnet located between DCH2 and DCH3 
\cite{Fanti:2007vi}. Each chamber has eight planes of sense wires, two 
horizontal, two vertical and two along each of two orthogonal $45^\circ$ 
directions. The spectrometer is located in a tank filled with helium at 
atmospheric pressure and separated from the decay volume by a thin
$Kevlar^{\textregistered}$ window with a thickess of 0.0031 
radiation lengths ($X_0$). A 16~cm diameter aluminium vacuum 
tube centred on the beam axis runs the length of the spectrometer through 
central holes in the Kevlar window, drift chambers and 
calorimeters. Charged particles are magnetically deflected in the horizontal
plane by an angle corresponding to a transverse momentum kick of 
\mbox{$120$ MeV/$c$}. The momentum resolution of the spectrometer is 
\mbox{$\sigma(p)/p = 1.02\% \oplus 0.044\%p$} ($p$ in \gevp), as derived from the
known properties of the spectrometer and checked with the measured invariant
mass resolution of $\kccc$ decays. The magnetic spectrometer is followed by
a scintillator hodoscope consisting of two planes segmented into horizontal
and vertical strips and arranged in four quadrants.

A liquid Krypton calorimeter (LKr) \cite{Barr:1995kp} is used to reconstruct 
$\pi^0 \rightarrow \gamma \gamma$ decays. It is an almost homogeneous 
ionization chamber with an active volume of 
$\sim 10 $~m$^3$ of liquid krypton, segmented transversally into 13248 
2~cm $\times$ 2~cm projective cells by a system of
Cu-Be ribbon electrodes, and with no longitudinal segmentation. 
The calorimeter is 27 $X_0$ thick and has an energy resolution 
$\sigma(E)/E = 0.032/\sqrt{E} \oplus 0.09/E \oplus 0.0042$ (E in \geve). 
The space resolution for single electromagnetic showers can be parameterized as
$\sigma_x = \sigma_y = 0.42/\sqrt{E} \oplus 0.06$~cm for each transverse 
coordinate $x,y$.

An additional hodoscope consisting of a plane of scintillating fibers is 
installed in the LKr calorimeter at a depth of $\sim 9.5 ~X_0$ with the
purpose of sampling electromagnetic showers. It is
divided into four quadrants, each consisting of eight bundles of vertical
fibers optically connected to photomultiplier tubes.

\section{Event selection and reconstruction}
\label{evselrec}

The $\kcnn$ decays are selected by a two level trigger. 
The first level requires a signal in at least one quadrant of the 
scintillator hodoscope (Q1) in coincidence
with the presence of energy depositions in LKr consistent with at least two
photons (NUT). At the second level (MBX), an on-line processor
receiving the drift chamber information reconstructs the momentum of
charged particles and calculates the missing mass under the assumption
that the particle is a $\pi^\pm$ originating from the decay of a 60~\gevp 
~$K^\pm$ travelling along the nominal beam axis.
The requirement that the missing mass is not consistent with the $\pi^0$ mass
rejects most of the main $K^\pm \rightarrow \pi^\pm \pi^0$ background. The
typical rate of this trigger is $\sim 15,000$ per burst.

Events with at least one charged particle track having a momentum above 
5~\gevp, measured  with a maximum error of 6\% (much larger than 
the magnetic spectrometer resolution),
and at least four energy clusters in the LKr, 
each consistent, in terms of size and 
energy, with the electromagnetic shower produced by a photon of energy above 3~\geve,
are selected for further analysis. 
In addition, the relative track and photon timings must be consistent with the 
same event within 10~ns, and the clusters must be in time between each
other within 5~ns.

The distance between any two photons in the LKr is required to be larger
than 10~cm, and the distance between each photon and the impact point of any
track on the LKr front face must exceed 15~cm. Fiducial cuts on the distance of each photon
from the LKr edges and centre are also applied in order to ensure 
full containment of the electromagnetic showers. In addition, because of the
presence of $\sim100$ LKr cells affected by readout problems (``dead cells''),
the minimum distance between the photon and the nearest LKr dead cell
is required to be at least 2~cm. 

At the following step of the analysis we check the consistency of the surviving
events with the $\kcnn$ decay hypothesis. We assume that each possible pair of
photons originates from a $\pi^0 \rightarrow \gamma \gamma$ decay and we 
calculate the distance $D_{ij}$ between the $\pi^0$ decay vertex and
the LKr front face:

$$
D_{ij} = 
\frac
{\sqrt{E_i E_j} R_{ij}}
{m_0}
$$   
where $E_i$,$E_j$ are the energies of the $i$-th and $j$-th photon, respectively,
$R_{ij}$ is the distance between their impact points on LKr, and $m_0$ is
the $\pi^0$ mass. 

Among all possible $\pi^0$ pairs, only those with  $D_{ij}$ values differing 
by less than 500~cm are retained further, and 
the distance $D$ of the $K^\pm$ decay vertex from the LKr is taken as the
arithmetic average of the two $D_{ij}$ values. This choice gives 
the best $\p0p0$ invariant mass resolution near threshold: at $\mmm = 2m_+$
it is \mbox{$\sim 0.56$ MeV/$c^2$}, increasing monotonically to \mbox{$\sim 1.4$ MeV/$c^2$} 
at the upper edge of the physical region. The reconstructed distance of
the decay vertex from the LKr is further required to be at least 2~m
downstream of the final beam collimator to exclude $\pi^0$-mesons
produced from beam particles interacting in the collimator material
(the downstream end of the final beam collimator is at $Z=-18$~m).

Because of the long decay volume, a photon emitted at small angle to the beam
axis may cross the aluminium vacuum tube in the spectrometer or 
the DCH1 central flange, and convert to $e^+ e^-$ before reaching the LKr. 
In such a case the photon must be rejected because its energy
cannot be measured precisely. To this purpose, for each photon detected in LKr
we require that its distance from the nominal beam axis at the DCH1 plane
must be $> 11$~cm, assuming an origin on axis at $D-400$~cm.
In this requirement we take into account the resolution of the $D$
measurement (the $rms$ of the
difference between $D$ values for the two photon pairs
distribution is about 180~cm).

Each surviving $\pi^0$ pair is then combined with a charged particle track,
assumed to be a $\pi^\pm$. Only those combinations with a total
$\pi^\pm \pi^0 \pi^0$ energy between $54$ and $66$~\geve, consistent with
the beam energy distribution, are retained, and the $\pi^\pm \pi^0 \pi^0$ 
invariant mass $M$ is calculated, after correcting the charged 
track momentum vector for the effect of the small measured residual magnetic
field in the decay volume (this correction uses the decay vertex position, D, 
as obtained from LKr information).

For each $\pi^\pm \pi^0 \pi^0$ combination, the  energy-weighed average coordinates 
(center-of-gravity, COG) $X_{COG},Y_{COG}$ are calculated 
at each DCH plane using the photon impact points on LKr and 
the track parameters measured before the magnet 
(so the event COG is a projection of the initial kaon line of flight).
Acceptance cuts are then applied on the COG radial position on each
DCH plane in order to select only $\kcnn$ decays
originating from the beam axis.\footnote{The beam is focused at the DCH1 
plane, where its width is $\sim 0.45$~cm.} 
In addition, we require a minimal separation between the 
COG and the charged track coordinates $X_{t}, Y_{t}$, as measured in
each DCH plane:
\begin{eqnarray}
\nonumber \sqrt{X_{COG}^2 + Y_{COG}^2} <  R^{COG}_{max}, \\
\nonumber \sqrt{(X_{COG} - X_t)^2 + (Y_{COG} - Y_t)^2} >  R^{COG-track}_{min}, 
\end{eqnarray}
where the limits depend on the COG and track impact point distributions
at each drift chamber (see Table \ref{rcuts}).
\begin{table}[ht]
\caption{Acceptance cuts on event COG and charged track coordinates.}
\label{rcuts}
\begin{center}
\begin{tabular}{|l|l|l|}                                                   \hline
Drift chamber       & $R^{COG}_{max}$ (cm) & $R^{COG-track}_{min}$ (cm) \\ \hline
DCH1                &           2.0        &           17.0             \\ 
DCH2                &           2.0        &           19.0             \\ 
DCH3                &           2.0        &           19.0             \\ 
DCH4                &           3.0        &           15.5             \\ \hline
\end{tabular}
\end{center}
\end{table}

The values of $R^{COG-track}_{min}$ take into account both the beam width
(the cut is made with respect to each event COG rather than to
the nominal beam center) and the area where the track impact point 
distribution is still sensitive to the detailed features of the beam shape. 
In this way the effect of these cuts does not depend strongly on the beam
shape and on the precise knowledge of the beam position in space
(during data taking, the average beam transverse position was observed 
to move slightly by up to 2~mm). This cut removes about 28\% of events, 
mainly at large $\mm2$, but the statistical precision of the final results 
on the $\pi \pi$ scattering lengths is not affected.

For events with more than one accepted track-cluster combination
($\sim1.8$\% of the total), the $\kcnn$ decay is selected
as the $\pi^\pm \pi^0 \pi^0$ combination minimizing a quality estimator 
based on two variables: the difference $\Delta D$
of the two $D_{ij}$ values and the difference  $\Delta M$
between the $\pi^\pm \pi^0 \pi^0$ invariant mass and the nominal $K^\pm$ mass
\cite{Amsler:2008zz}:

$$\left(\frac{\Delta D}{rms_D(D)}\right)^2 + \left(\frac{\Delta M}{rms_M(D)}\right)^2,$$ 
where the space and mass resolutions $rms_D, rms_M$ are functions of 
$D$, as obtained from the measured $\Delta D$ and $\Delta M$ distributions.

\begin{figure}[ht]

\begin{center}
\resizebox{1.05\columnwidth}{!}{%
\setlength{\unitlength}{1mm}

\begin{picture}(100.,100.)                      

\includegraphics[width=100mm]{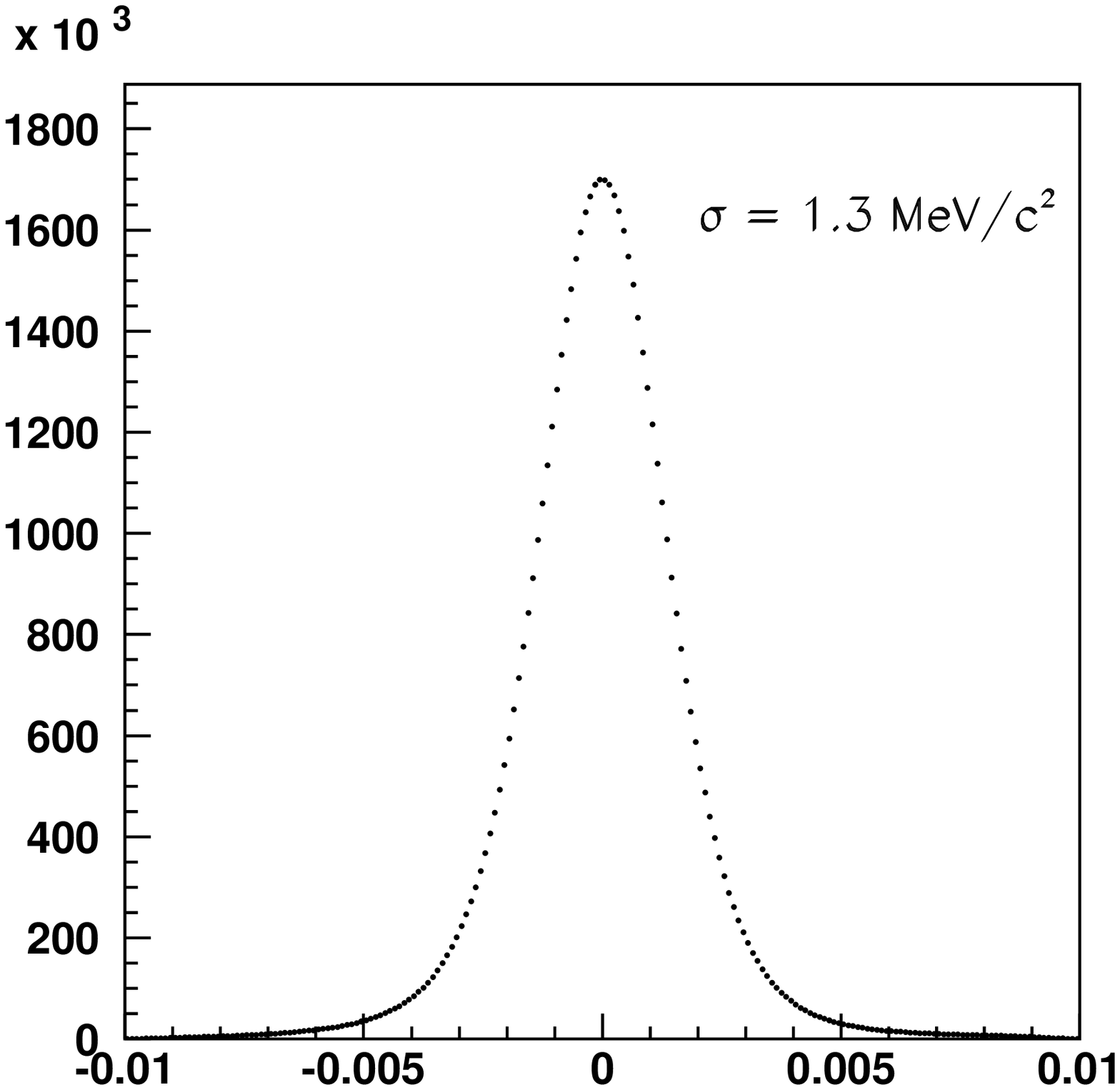}  
\put(-50.,2.){\makebox(0,0){\large $\Delta M$ (\gevm)}}
\put(-99.,50.){\makebox(0,0){\rotatebox{90}{events/0.0001  \gevm}}}
 
\end{picture}
}
\end{center}

\caption{Distribution of the difference between the $\pi^\pm \pi^0 \pi^0$
invariant mass and the nominal $K^\pm$ mass for the selected
$\kcnn$ decays.}
\label{kmass}

\end{figure}

Fig.~\ref{kmass} shows the distribution of $\Delta M$,
the difference between the $\pi^\pm \pi^0 \pi^0$  invariant mass
and the nominal $K^\pm$ mass for the selected $\kcnn$ decays (a
total of $6.031\times 10^7$ events). This distribution is dominated by the gaussian
$K^\pm$ peak, with a resolution \mbox{$\sigma = 1.3$ MeV/$c^2$}. There 
are small non Gaussian tails originating from unidentified 
$\pi^\pm \rightarrow \mu^\pm$ decay in flight or wrong photon pairing. 
The fraction of events with wrong photon pairing in this sample is $0.19 \%$, 
as estimated by the Monte Carlo simulation described in the next Section.

\begin{figure}[ht]

\begin{center}
\resizebox{1.1\columnwidth}{!}{%
\setlength{\unitlength}{1mm}

\begin{picture}(100.,100.)                       

\includegraphics[width=100mm]{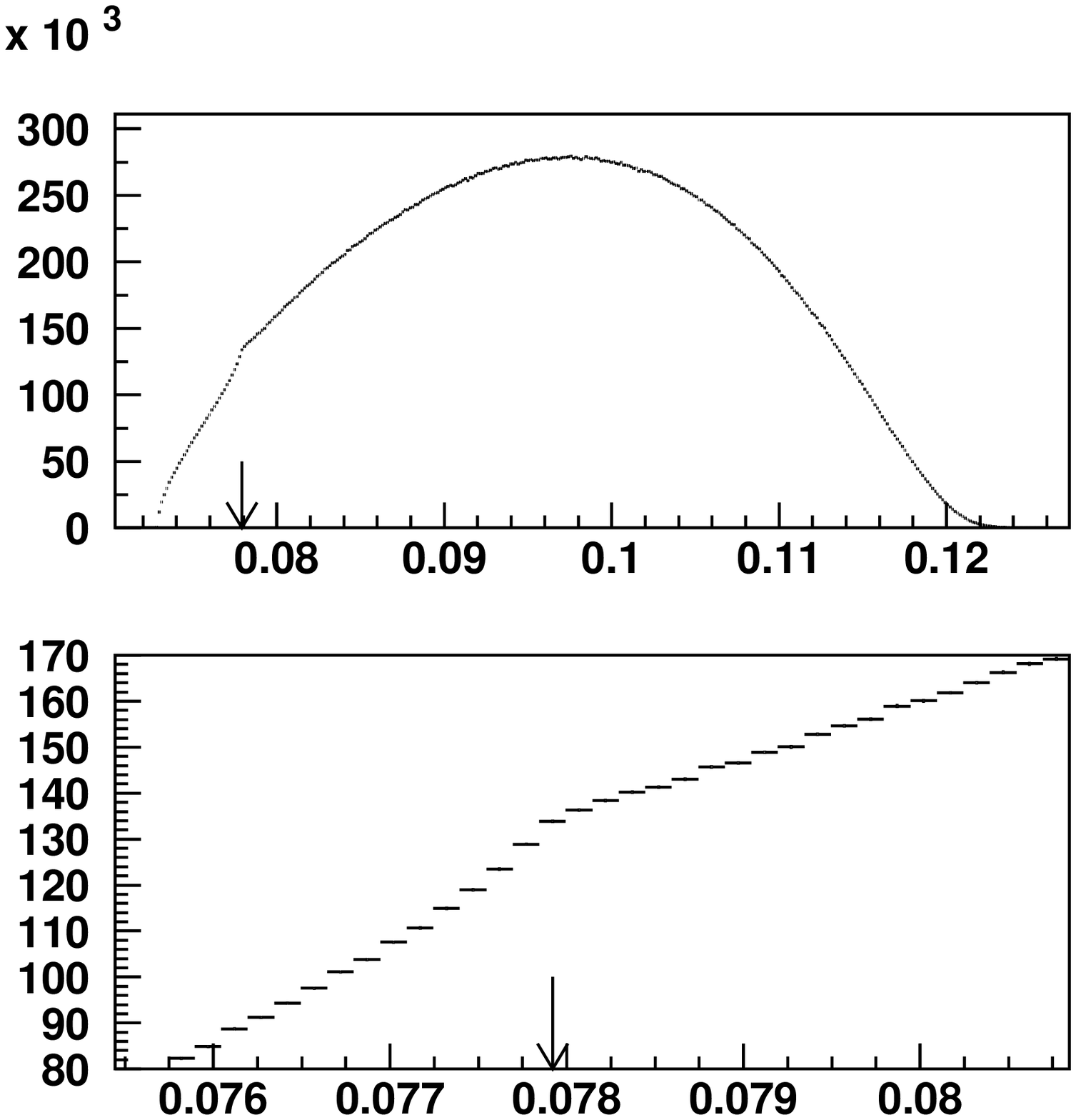}   
\put(-80.,80.){\makebox(0,0){\bf a}}
\put(-80.,35.){\makebox(0,0){\bf b}}
\put(-50.,2.){\makebox(0,0){\large $\mm2$  \gevmsq}}
\put(-97.,50.){\makebox(0,0){\rotatebox{90}{events/0.00015  \gevmsq}}}
\end{picture}
}
\end{center}

\caption{{\bf a}: distribution of $\mm2$, the square of the $\p0p0$ invariant
mass; {\bf b}: enlargement of a narrow region centred at $\mm2 = (2m_+)^2$
(this point is indicated by the arrow). The statistical error bars are
also shown in these plots.}
\label{m00sq}

\end{figure}

Fig. \ref{m00sq} shows the distribution of the square of the $\p0p0$ 
invariant mass, $\mm2$, for the final event sample. This distribution is 
displayed with a bin width of 0.00015~\gevmsq, with the $51^{st}$
bin centred at $\mm2 = (2m_+)^2$ (for most of the physical region the bin 
width is smaller than the $\mm2$ resolution, which is 0.00031~\gevmsq~at 
$\mm2 = (2m_+)^2$). The cusp at $\mm2 = (2m_+)^2 = 0.07792$~\gevmsq~is 
clearly visible.
 
\section{Monte Carlo simulation}
\label{mcsimul}
Samples of simulated  $\kcnn$ events $\sim10$ times larger than the data have
been generated using a full detector simulation based on the GEANT-3 package
\cite{Brun:1978fy}. This Monte Carlo (MC) program takes into account
all detector effects, including the trigger efficiency and the
presence of a small number $(<1\%)$ of ``dead'' LKr cells. It also
includes the simulation of the beam line; the beam parameters are
tuned for each SPS burst using fully
reconstructed $\kccc$ events, which provide precise information on the average
beam angles and positions with respect to the nominal beam axis. Furthermore,
the requirement that the average reconstructed $\pi^\pm \pi^+ \pi^-$ invariant mass
is equal to the nominal $K^\pm$ mass for both $K^+$ and $K^-$ fixes the
absolute momentum scale of the magnetic spectrometer for each charge sign and
magnet polarity, and monitors continuously the beam momentum distributions
during data taking. 

The Dalitz plot distribution of $\kcnn$ decays has been generated according to
a series expansion in the Lorentz-invariant variable $u = (s_3-s_0)/m_+^2$,  
where $s_i = (P_K - P_i)^2$ ($i$=1,2,3), 
$s_0 = (s_1+s_2+s_3)/3$, $P_K$ $(P_i)$
is the $K(\pi)$ four-momentum, and $i=3$ corresponds to the
$\pi^\pm$ \cite{Amsler:2008zz}. In our case $s_3 = \mm2$, and
$s_0 = (m_K^2 + 2m_0^2+m_+^2)/3$. For any given value of the generated
$\p0p0$ invariant mass the simulation provides the detection probability and
the distribution function for the reconstructed value of $\mm2$. This allows
the transformation of any theoretical distribution into an expected 
distribution which can be compared directly with the measured one.

\section{ \boldmath Determination of the $\pi\pi$ scattering lengths
  $a_0$ and $a_2$  \unboldmath}
\label{sclen}
The sudden change of slope (``cusp'' ) observed in the $\mm2$ distribution at 
$\mm2 = (2m_+)^2$ (see Fig. \ref{m00sq}) can be interpreted
\cite{Budini:1961} \cite{Cabibbo:2004gq} as a threshold effect from the decay
$\kccc$ contributing to the $\kcnn$ amplitude through the charge exchange
reaction $\pi^+ \pi^- \rightarrow \pi^0 \pi^0$. In the formulation
by Cabibbo \cite{Cabibbo:2004gq} the $\kcnn$ decay amplitude is described 
as the sum of two terms:
\begin{equation} 
{\cal M}(\kcnn) = {\cal M}_0 + {\cal M}_1,  
\end{equation}
where ${\cal M}_0$ is the tree level $\kcnn$ weak decay amplitude,
and ${\cal M}_1$ is the contribution from the $\kccc$ decay amplitude through
$\pi^+ \pi^- \rightarrow \pi^0 \pi^0$
charge exchange, with the normalization condition  ${\cal M}_1=0$ 
at $\mm2 = (2m_+)^2$. The contribution ${\cal M}_1$ is given by 
\begin{equation}
{\cal M}_1 = - 2 a_x m_+ {\cal M}_+\sqrt{\left(\frac{2m_+}{M_{00}}\right)^2-1},
\end{equation}
where $a_x$ is the S-wave $\pi^+ \pi^-$ charge exchange scattering length
(threshold amplitude), and
${\cal M}_+$ is the $\kccc$ decay amplitude at $M_{00}=2m_+$.
${\cal M}_1$ changes from real to imaginary at 
$M_{00} = 2m_+$ with the consequence that ${\cal M}_1$ interferes 
destructively with  ${\cal M}_0$ in the region $M_{00} < 2m_+$, while it
adds quadratically above it. In the limit of exact isospin symmetry
$a_x = (a_0-a_2)/3$, where $a_0$ and $a_2$ are the S-wave $\pi\pi$ 
scattering lengths in the $I=0$ and $I=2$ states, respectively.

However, it was shown in ref. \cite{Batley:2005ax} that a fit of this
simple formulation to the NA48/2 $\mm2$ distribution in the interval 
$0.074<\mm2<0.097$~\gevmsq
~using $a_x m_+$ as a free parameter gave only a qualitative description of
the data, with all data points lying systematically above the fit in the
region near  $\mm2 = (2m_+)^2$. It was also shown in
ref. \cite{Batley:2005ax} that a good fit could be obtained using a
more complete formulation of $\pi\pi$ final state interaction 
\cite{Cabibbo:2005ez} which took into account all rescattering processes at the
one-loop and two-loop level. 

In the following sections we present the determination of the $\pi\pi$ 
scattering lengths $a_0$ and $a_2$ by fits of the full data set
described in Section \ref{evselrec} to two theoretical approaches: the 
Cabibbo-Isidori (CI) formulation \cite{Cabibbo:2005ez}, and the more recent 
Bern-Bonn (BB) formulation \cite{Colangelo:2006va}.   

In the CI approach, the structure of the cusp singularity is treated
using unitarity, analiticity and cluster decomposition properties of
the $S$-matrix. The decay amplitude is expanded in powers of $\pi\pi$
scattering lengths up to order $(scattering~length)^2$, and electromagnetic
effects are omitted.

\begin{sloppypar}
The BB approach uses a non-relativistic Lagrangian framework, which
automatically satisfies unitarity and analiticity constraints, and
allows one to include electromagnetic contributions in a standard way
\cite{Bissegger:2008ff}.
\end{sloppypar}

In all fits we also need information on the $\kccc$ decay amplitude. To
this purpose, we use a sample of $4.709 \times 10^8$ $\kccc$ decays 
which are also measured in this experiment \cite{Batley:2007md}.

\subsection{Fits using the Cabibbo-Isidori theoretical formulation}
\label{ci}
In the Cabibbo-Isidori (CI) formulation \cite{Cabibbo:2005ez} the weak
amplitudes for $\kcnn$ and $\kccc$ decay at tree level are written as

\begin{equation}
{\cal M}_0 = 1 + \frac{1}{2}g_0u + \frac{1}{2}h_0u^2 + \frac{1}{2}k_0v^2, 
\label {amp0}
\end{equation}

\begin{equation}
{\cal M}_+ = A_+(1 + \frac{1}{2}gu + \frac{1}{2}hu^2 + \frac{1}{2}kv^2),
\label {amp+}
\end{equation}
respectively. In Eq.~(\ref{amp0}) $u=(s_3-s_0)/m_+^2$, where 
$s_0=(m_K^2+2m_0^2+m_+^2)/3$, while in Eq.~(\ref{amp+}) $u=(s_3-s_+)/m_+$, where
$s_+=m_K^2/3+m_+^2$; for both amplitudes $s_i=(P_K-P_i)^2$, where $P_K$ 
($P_i$) is the $K$ ($\pi$) four-momentum and $i=3$ corresponds to the
odd pion ($\pi^{\pm}$ from $\kcnn$, $\pi^{\mp}$ from $\kccc$ decay),
and $v = (s_1 - s_2)/m_+^2$. It must be noted that in ref.
\cite{Cabibbo:2005ez} the $v$ dependence of both amplitudes had been
ignored because the coefficients $k_0$ and $k$ were
consistent with zero from previous experiments. Within the very high 
statistical precision of the present experiment this assumption is no 
longer valid.

\begin{sloppypar}
Pion-pion rescattering effects are evaluated by means of an expansion
in powers of the $\pi \pi$ scattering lengths around the cusp point,
$\mm2 = (2m_+)^2$. The terms added to the
tree-level decay matrix elements depend on
five S-wave scattering lengths which are denoted by 
$a_x$, $a_{++}$, $a_{+-}$, $a_{+0}$, $a_{00}$, and describe 
$
\pi^+ \pi^- \rightarrow \pi^0 \pi^0,
 \pi^+ \pi^+ \rightarrow \pi^+ \pi^+,
 \pi^+ \pi^- \rightarrow \pi^+ \pi^-,
 \pi^+ \pi^0 \rightarrow \pi^+ \pi^0,
 \pi^0 \pi^0 \rightarrow \pi^0 \pi^0
$ 
scattering, respectively. In the limit of exact 
isospin symmetry these scattering lengths can all be expressed 
as linear combinations of $a_0$ and $a_{2}$. 
\end{sloppypar}

At tree level, omitting one-photon exchange diagrams, isospin symmetry 
breaking contributions to the elastic $\pi\pi$ scattering amplitude
can be expressed as a function of one parameter 
$ \eta = (m_+^2-m_0^2)/m_+^2 = 0.065$ 
\cite{VanKolck:1993ee,Maltman:1996nw,Knecht:1997jw}.
In particular, the ratio between the threshold amplitudes
$a_x$, $a_{++}$, $a_{+-}$, $a_{+0}$, $a_{00}$ and the corresponding
isospin symmetric amplitudes -- evaluated at the $\pi^\pm$ mass -- is
equal to
$1-\eta $ for 
$
\pi^+ \pi^+ \rightarrow \pi^+ \pi^+,
 \pi^+ \pi^0 \rightarrow \pi^+ \pi^0,
 \pi^0 \pi^0 \rightarrow \pi^0 \pi^0
$,
$1+\eta $ for 
$\pi^+ \pi^- \rightarrow \pi^+ \pi^-$,
and 
 $1+\eta/3 $ for 
$\pi^+ \pi^- \rightarrow \pi^0 \pi^0$.
These corrections have been applied in order to extract $a_0$ and $a_2$ from
the fit to the $\mm2$ distribution.

The CI formulation \cite{Cabibbo:2005ez} includes all one-loop
and two-loop rescattering diagrams and can be used to fit both $\kcnn$
and $\kccc$ decay distributions. However, rescattering effects are much smaller 
in $\kccc$ than in the $\kcnn$ decay because the invariant mass of any two-pion 
pair is always  $\ge 2m_+$. Indeed, a good fit to the $\kccc$ Dalitz plot
\cite{Batley:2007md} can be obtained
with or without the addition of rescattering terms to the tree-level
weak amplitude of $\kccc$ decay. We have checked
that both the values of the best fit parameters and their statistical errors,
as obtained from fits to the $\mm2$ distribution of   
$\kcnn$ decay,  
undergo negligible changes whether or not rescattering effects are
included in the $\kccc$ decay amplitude. This can be understood from
the fact that the $\kccc$ decay amplitude enters
into the CI formulation of rescattering effects in $\kcnn$ decays as
the complete expression given by Eq.~(\ref{amp+}). Thus Eq.~(\ref{amp+}),
with parameters extracted from a fit to the $\kccc$ data, provides an
adequate phenomenological description of $\kccc$ decay which can be
used in calculating rescattering effects in $\kcnn$ decay.

In the fits to the $\mm2$ distribution from $\kcnn$ decay, the free parameters
are $(a_0-a_2)m_+$, $a_2m_+$, $g_0$, $h_0$,
and an overall normalization constant. The coefficient $k_0$ cannot be
directly obtained from a fit to the  $\mm2$ distribution. Its value is
determined independently from the Dalitz plot distribution of $\kcnn$ decays,
as described in the Appendix. The value $k_0 = 0.0099$ is kept fixed
in the fits.

All ${\cal M}_+$ parameters are fixed from data: the coefficients
$g$, $h$, $k$ are obtained from a separate fit to the $\kccc$ decay Dalitz plot 
\cite{Batley:2007md}, using ${\cal M}_+$ as given by
Eq.~(\ref{amp+}), and taking into account Coulomb effects; and $A_+$ is obtained
from  the measured ratio, $R$, of the $\kccc$ and $\kcnn$ decay rates,
$R = 3.175 \pm 0.050$ \cite{Amsler:2008zz}, which is proportional
to $A_+^2$. The fit gives $g = -0.2112 \pm 0.0002$, $h = 0.0067 \pm 0.0003$, 
$k = -0.00477 \pm 0.00008$; and we obtain $A_+ = 1.925 \pm 0.015$. These values
are kept fixed in the fits to the $\mm2$ distribution from $\kcnn$ decay.     
 
\begin{figure}[h]

\begin{center}
\resizebox{1.05\columnwidth}{!}{%
\setlength{\unitlength}{1mm}

\begin{picture}(100.,100.)                 

\includegraphics[width=100mm]{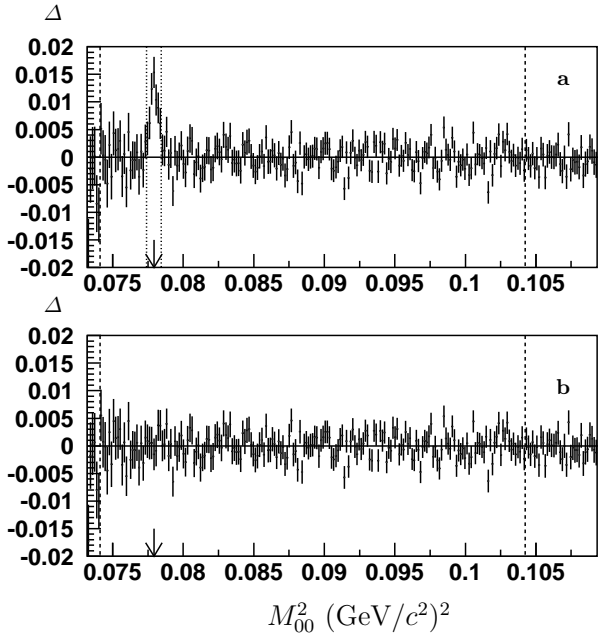} 
\put(-90.,90.){\makebox(0,0){$\Delta$}}
\put(-90.,47.){\makebox(0,0){$\Delta$}}
\put(-15.,80.){\makebox(0,0){\bf a}}
\put(-15.,35.){\makebox(0,0){\bf b}}
\put(-45.,1.){\makebox(0,0){\large $\mm2$ \gevmsq}}
 
\end{picture}
}
\end{center}

\caption{$\Delta =$ (data -- fit)/data versus $\mm2$ for the
rescattering formulation of ref. \cite{Cabibbo:2005ez}: {\bf a} -- fit with no
pionium formation and excluding seven consecutive bins centred at 
$\mm2 = (2m_+)^2$ (the excluded region is shown by the two vertical
dotted lines; {\bf b} -- fit with pionium $\fitA$ (see text). The two vertical dashed lines show
the $\mm2$ interval used in the fit. The point $\mm2 = (2m_+)^2$  is 
indicated by the arrow.}
\label{cifits}

\end{figure}

As explained in Section \ref{systematics} all fits are performed over the
$\mm2$ interval from $0.074094 $ to $0.104244$~\gevmsq ~(bin 26 to 226).
The CI formulation \cite{Cabibbo:2005ez} does not include
radiative corrections, which are particularly important near $M_{00}=2m_+$,
and contribute to the formation of $\pi^+\pi^-$ atoms (``pionium'').
For this reason we first exclude from the fit a group of seven consecutive bins
centred at $M_{00}^2=4m_+^2$ (an interval of $\pm 0.94$ MeV/$c^2$ in $M_{00}$). 
The quality of this fit is illustrated in Fig. \ref{cifits}a, which
displays the quantity
$\Delta \equiv $ (data -- fit)/data as a function of $\mm2$. 
The small excess of events from pionium formation is clearly visible.

Pionium formation and its dominating decay to $\pi^0\pi^0$ are taken into 
account in the fit by multiplying the
content of the bin centred at $\mm2 = 4m_{+}^{2}$ (bin 51) by
$1+f_{atom}$, where $1+f_{atom}$ describes the contribution from
pionium formation and decay. The pionium width is much narrower than the 
bin width, since its mean lifetime is measured to be $\sim 3 \times 10^{-15}$~s
\cite{Adeva:2005pg}; however, the $\mm2$ resolution is taken into
account in the fits as described in the last paragraph of Section 3. 
The results of a fit with $f_{atom}$ as a free parameter and with no
excluded bins near $M_{00}^2=4m_+^2$ are given in Tables \ref{ftab1}
and \ref{ftab2} (fit $\fitA$): the quality of this fit is shown in 
Fig.~\ref{cifits}b. The best fit value $f_{atom} = 0.0533 \pm 0.0091$ 
corresponds to a rate of $K^{\pm} \rightarrow \pi^{\pm} +$ pionium decay,
normalized to the $\kccc$ decay rate, of $(1.69 \pm 0.29) \times 10^{-5}$,
which is larger than the predicted value $\sim 0.8\times 10^{-5}$
\cite{Pilkuhn:1978rf,Silagadze:1994wr}.
As discussed in Section \ref{radiative}, this difference is due to additional
radiative effects, which are not taken into account in the CI formulation
\cite{Cabibbo:2005ez} and, contrary to pionium formation and
decay, affect more than one bin. For this reason  
for the fits without the radiative effects taken into account        
we prefer to fix
$f_{atom} = 0.0533$ and to exclude from the fit the seven consecutive bins
centred at $M_{00}^2=4m_+^2$. The results of this fit are listed as Fit $\fitB$ in
Tables \ref{ftab1} and \ref{ftab2}.

We have also performed fits using the constraint between $a_2$
and $a_0$ predicted by analyticity and chiral symmetry \cite{Colangelo:2001sp}
(we refer to this constraint as the ChPT constraint):
   
\begin{eqnarray}
\nonumber
a_2m_+ = (-0.0444\pm 0.0008) + 0.236(a_0m_+ -0.22) \\
-0.61(a_0m_+ -0.22)^2 -9.9(a_0m_+ -0.22)^3
\label {ChPT}
\end{eqnarray}  

The results of these fits are shown in Tables \ref{ftab1} and
\ref{ftab2} (fits $\fitC$ and $\fitD$). For fit $\fitC$ no
bins near the cusp point are excluded and $f_{atom}$ is a free
parameter, while for fit $\fitD$ the seven bins centred at $M_{00}^2=4m_+^2$ are
excluded and $f_{atom}$ is kept fixed at the value obtained from fit $\fitC$.

\subsection{Fits using the Bern-Bonn theoretical formulation}
\label{bb}

The Bern-Bonn (BB) formulation \cite{Colangelo:2006va} describes the
$K \rightarrow 3\pi$ decay amplitudes using two expansion parameters: $a$, the
generic $\pi \pi$ scattering amplitude at threshold; and a formal
parameter $\epsilon$ such that in the $K$-meson rest frame the pion
momentum is of order $\epsilon$, and its kinetic energy $T$ is of order
$\epsilon^2$. In the formulation of ref. \cite{Colangelo:2006va} the 
$K \rightarrow 3\pi$ decay amplitudes include terms up to
$O(\epsilon^2, a \epsilon^3, a^2 \epsilon^2)$. However, in the
formulae used in the fits described below these amplitudes
include terms up to $O(\epsilon^4, a \epsilon^5, a^2 \epsilon^2 )$. In
the BB formulation the description of the 
$K \rightarrow 3\pi$ decay amplitudes is valid over the full physical region
\footnote{We thank the Bern-Bonn group for providing the computer code
which calculates the $K \rightarrow 3\pi$ decay amplitudes.}.

At tree level the $K \rightarrow 3\pi$ decay amplitudes are expressed as polynomials
containing terms in $T_3$, $T_3^2$, and $(T_1-T_2)^2$, where $T_3$ is
the kinetic energy of the ``odd'' pion ($\pi^{\pm}$ from $\kcnn$, 
$\pi^{\mp}$ from $\kccc$ decay) in the $K^{\pm}$ rest frame,
while $T_1$ and $T_2$ are the kinetic energies of the two same-sign pions. 
Since these variables can be expressed as functions of the relativistic 
invariants $u$ and $v$ defined previously, for consistency with the fits
described in the previous subsection we prefer to use the same forms
as given in Eqs. (\ref{amp0}) and (\ref{amp+}). It must be noted, however,
that the best fit polynomial coefficients are not expected to be equal
to those obtained from the fits to the CI formulation
\cite{Cabibbo:2005ez} because the loop diagram contributions are
different in the two formulations.

As for CI, also in the BB formulation rescattering effects are much smaller in 
$\kccc$ than in the $\kcnn$ decay, and a good fit to the $M_{\pm \pm}^2$ 
distribution alone can be obtained
with or without the addition of rescattering terms to the tree-level
weak amplitude of $\kccc$ decay. However, contrary to CI, the coefficients
of the tree-level $\kccc$ amplitudes enter into the $\kcnn$ rescattering terms
in different combinations. Therefore, the use of a phenomenological
description of the $\kccc$ decay amplitude extracted from a fit to
$\kccc$ data alone is not justified in this case. Thus,
in order to obtain a precision on the fit parameters which matches
the BB approximation level, the value of each coefficient of the 
$\kccc$ tree-level amplitude is obtained from  
the fit.\footnote{Nevertheless, if one fixes the coefficients $g,h,k$
in the fit to the values obtained from fits to $\kccc$ data only
with or without rescattering terms, the corresponding variations of
the best fit $a_0,a_2$ values are much smaller than the $a_0,a_2$ 
statistical errors.}

We perform simultaneous fits to two distributions: the $\mm2$
distribution described in Section \ref{evselrec} and the
$M_{\pm \pm}^2$ distribution from $\kccc$ decay, obtained as a
projection of the Dalitz plot described in
ref. \cite{Batley:2007md}. This latter distribution is made with 
the same binning as for the $\mm2$ distribution from $\kcnn$ decay
and consists of $4.709 \times 10^8$ events.

\begin{sloppypar}
All fits are performed over the $\mm2$ interval from 
$0.074094 $ to $0.104244$~\gevmsq ~(bin 26 to 226), and from
$0.080694 $ to $0.119844$~\gevmsq ~(bin 70 to 330) for the
$M_{\pm \pm}^2$ distribution from $\kccc$ decay. As
for the $\mm2$ distribution from $\kcnn$ decay, a very large sample of
simulated $\kccc$ decays (see ref. \cite{Batley:2007md}) is used to obtain
the detection probability and the distribution function for the reconstructed
value $M_{\pm \pm}^2$ for any generated value of $M_{\pm \pm}^2$. 
\end{sloppypar}

In all fits the free parameters are $(a_0-a_2)m_+$ and $a_2m_+$ (or only
$a_0m_+$ for the fit using the ChPT constraint given by Eq.~(\ref{ChPT})), 
the coefficients of the tree-level weak
amplitudes $g_0$, $h_0$, $g$, $h$, $k$ (see Eqs.~(\ref{amp0}, \ref{amp+})),
and two overall normalization constants (one for each distribution).
The coefficient $k_0$ (see Eq.~(\ref{amp0})) is determined independently
from a separate fit to the Dalitz plot distribution of $\kcnn$ decays 
(see the Appendix
). The fixed value $k_0 = 0.0085$ 
is used in the fits. In some of the fits the contribution from
pionium formation, described by $f_{atom}$, is also a free parameter. 
 
Since the detection of $\kcnn$ and $\kccc$ decays involves different detector
components and different triggers (no use of LKr information is made to
select $\kccc$ decays), the ratio of the detection efficiencies for
the two decay modes is not known with the precision needed to extract
the value of $A_+$ (see Eq.~(\ref{amp+})) from the fit. Therefore, as
for the CI fits, also for the BB fits $A_+$ is obtained from  the ratio of the 
$\kccc$ and $\kcnn$ decay rates, measured by other experiments, 
$R = 3.175 \pm 0.050$ \cite{Amsler:2008zz}.

Tables \ref{ftab1} and \ref{ftab2} show the results of a fit
(fit $\fitE$) using $f_{atom}$ as a free
parameter and including all bins around the cusp point in the fit; for
fit $\fitF$ the value of $f_{atom}$ is fixed and seven bins centred at 
$\mm2=4m_+^2$ are excluded. A comparison with the results of the
corresponding CI fits
(fits $\fitA$ and $\fitB$, respectively) shows that the difference between the
best fit values of $(a_0-a_2)m_+$ is rather small (about 3\%),
while the difference between the two $a_2m_+$ values is much
larger. We note that in the BB fits $a_2m_+$ has a stronger
correlation with other fit parameters than in the CI fits
(see Tables \ref{cabicorr} and \ref{berncorr}).

\begin{table*}
\caption{Fit results without radiative corrections: $\pi \pi$ scattering 
parameters. Parameter values without errors have been kept fixed in the fit or
calculated using the constraint between $a_2$ and $a_0$ given by Eq.~(\ref{ChPT}).
}
\begin{center} 
\begin{tabular}{|l|l|l|l|l|l|} \hline
Fit & $\chi^2/NDF$ & $a_0 m_+$ & $a_2 m_+$ & $(a_0-a_2) m_+$ & $f_{atom}$ \\ \hline
$ \fitA $ &  206.3/195 &  $  0.2334$(48) & $ -0.0392$(80) & $  0.2727$(46) & $  0.0533$(91) \\
$ \fitB $ &  201.6/189 &  $  0.2345$(50) & $ -0.0344$(86) & $  0.2689$(50) & $  0.0533$ \\
$ \fitC $ &  210.6/196 &  $  0.2336$(27) & $ -0.0413$ & $  0.2749$(21) & $  0.0441$(76) \\
$ \fitD $ &  207.6/190 &  $  0.2326$(27) & $ -0.0415$ & $  0.2741$(21) & $  0.0441$ \\
$ \fitE $ &  462.9/452 &  $  0.2122$(107) & $ -0.0693$(136) & $  0.2815$(43) & $  0.0530$(95) \\
$ \fitF $ &  458.5/446 &  $  0.2182$(109) & $ -0.0594$(143) & $  0.2776$(48) & $  0.0530$ \\
$ \fitG $ &  467.3/453 &  $  0.2321$(33) & $ -0.0417$ & $  0.2737$(26) & $  0.0647$(76) \\
$ \fitH $ &  459.8/447 &  $  0.2301$(34) & $ -0.0421$ & $  0.2722$(27) & $  0.0647$ \\
\hline \end{tabular}
\end{center}

\label{ftab1}
\end{table*}

\begin{table*}
\caption{Fit results without radiative corrections: coefficients of the 
tree-level $K \rightarrow 3\pi$  weak decay amplitudes.
Parameter values without errors have been kept fixed in the fit.}
\begin{center} 
\begin{tabular}{|l|l|l|l|l|l|l|} \hline
Fit & $g_0$ & $h_0$ & $k_0$ & $g$ & $h$ & $k$ \\ \hline
$ \fitA $ &  $  0.6512$(19) & $ -0.0386$(23) & $  0.0099$ & $ -0.2112$ & $  0.0067$ & $ -0.0048$ \\
$ \fitB $ &  $  0.6502$(20) & $ -0.0375$(23) & $  0.0099$ & $ -0.2112$ & $  0.0067$ & $ -0.0048$ \\
$ \fitC $ &  $  0.6485$(9) & $ -0.0436$(8) & $  0.0099$ & $ -0.2112$ & $  0.0067$ & $ -0.0048$ \\
$ \fitD $ &  $  0.6485$(9) & $ -0.0438$(8) & $  0.0099$ & $ -0.2112$ & $  0.0067$ & $ -0.0048$ \\
$ \fitE $ &  $  0.6117$(49) & $ -0.0589$(56) & $  0.0085$ & $ -0.1793$(20) & $ -0.0015$(20) & $ -0.0053$(23) \\
$ \fitF $ &  $  0.6154$(51) & $ -0.0550$(57) & $  0.0085$ & $ -0.1811$(23) & $ -0.0012$(20) & $ -0.0059$(22) \\
$ \fitG $ &  $  0.6215$(10) & $ -0.0480$(9) & $  0.0085$ & $ -0.1837$(5) & $ -0.0011$(20) & $ -0.0074$(20) \\
$ \fitH $ &  $  0.6215$(10) & $ -0.0483$(9) & $  0.0085$ & $ -0.1840$(5) & $ -0.0008$(20) & $ -0.0071$(20) \\
\hline \end{tabular}
\end{center}

\label{ftab2}
\end{table*}

\begin{table}[ht]
\caption{Parameter correlations for the CI fits (fit $\fitB$ in Table \ref{ftab1}).}
\begin{tabular}{|l|rrrr|} \hline
       & $g_0$ & $h_0$ & $a_0-a_2$ & $a_2$  \\ \hline 
$g_0$  & $ 1.000$ &  &  &   \\
$h_0$  & $-0.701$ & $ 1.000$ &  &   \\
$a_0-a_2$  & $ 0.777$ & $-0.793$ & $ 1.000$ &   \\
$a_2$  & $-0.902$ & $ 0.936$ & $-0.869$ & $ 1.000$  \\
\hline 
\end{tabular}

\label{cabicorr}
\end{table}

\begin{table}[ht]
\caption{Parameter correlations for the BB fits (fit $\fitF$ in Table \ref{ftab1}).}
\resizebox{\columnwidth}{!}{%
\begin{tabular}{|l|rrrrrrr|} \hline
       & $g_0$ & $h_0$ & $ g$ & $ h$ & $ k$ & $a_0-a_2$ & $a_2$  \\ \hline 
$g_0$  & $ 1.000$ &  &  &  &  &  &   \\
$h_0$  & $ 0.996$ & $ 1.000$ &  &  &  &  &   \\
$ g$  & $-0.970$ & $-0.960$ & $ 1.000$ &  &  &  &   \\
$ h$  & $ 0.206$ & $ 0.181$ & $-0.247$ & $ 1.000$ &  &  &   \\
$ k$  & $-0.399$ & $-0.423$ & $ 0.359$ & $ 0.803$ & $ 1.000$ &  &   \\
$a_0-a_2$  & $-0.853$ & $-0.817$ & $ 0.932$ & $-0.402$ & $ 0.141$ & $ 1.000$ &   \\
$a_2$  & $ 0.976$ & $ 0.987$ & $-0.958$ & $ 0.099$ & $-0.503$ & $-0.794$ & $ 1.000$  \\
\hline 
\end{tabular}

}
\label{berncorr}
\end{table}

Fits $\fitG$ and $\fitH$ (see Tables \ref{ftab1} and \ref{ftab2}) are similar
to $\fitE$ and $\fitF$, respectively, but the ChPT constraint given by
Eq.~(\ref{ChPT}) is used. Here the best fit value of $a_0m_+$ agrees
well with the value obtained from the CI fit (fit $\fitD$).

\section{Radiative effects}
\label{radiative}

\subsection{Radiative correction outside the cusp point}

Radiative corrections to both $\kcnn$ and $\kccc$ decay channels
have been recently studied by extending the BB formulation 
\cite{Colangelo:2006va} to include real and virtual photons 
\cite{Bissegger:2008ff}. In the $K^{\pm}$ rest frame the emission of 
real photons is allowed only for photon energies $E < E_{cut}$.

We have performed simultaneous fits to the $\mm2$ distribution from
$\kcnn$ and to the $M_{\pm \pm}^2$ distribution from $\kccc$ decays
using the formulation of ref. \cite{Bissegger:2008ff}. Our event selection
does not exclude the presence of additional photons; however,
energetic photons emitted in $K^{\pm}$ decays result
in a reconstructed $\pi^{\pm} \pi^0 \pi^0$ invariant mass lower than
the $K$ mass. We set $E_{cut}$ = 0.010~\geve ~in order to be consistent
with the measured $\pi^{\pm} \pi^0 \pi^0$ invariant mass distribution
shown in Fig. \ref{kmass} (the same is true for the 
$\pi^{\pm} \pi^+ \pi^-$ invariant mass distribution from $\kccc$ 
decay measured in this experiment \cite{Batley:2007md}).
For each fit we adjust the value of $A_+$ (see Eq. (\ref{amp+})) so that
the ratio of the $\kccc$ and $\kcnn$ decay rates is consistent with
the measured one \cite{Amsler:2008zz}.

The formulation of ref. \cite{Bissegger:2008ff} does not include pionium
formation, and the $\kcnn$ amplitude, $A_{00+}^{rad}$, has a non-physical
singularity at $\mm2 = (2m_+)^2$. To avoid problems in the fits,  the square of
decay amplitude at the center of bin 51, where the singularity occurs, is
replaced by $|A_{00+}|^2 (1+f_{atom})$, where $A_{00+}$ is the decay amplitude
of the BB formulation without radiative corrections \cite{Colangelo:2006va},
and $f_{atom}$ is again a free parameter.

The results of simultaneous fits to the $\mm2$ distribution from
$\kcnn$ decays, and to the $M_{\pm \pm}^2$ distribution from $\kccc$ decay are 
shown in Tables \ref{femtab1} and \ref{femtab2}. In all these fits
the $\mm2$ and $M_{\pm \pm}^2$ intervals are equal to those of the fits
described in Sections \ref{ci} and \ref{bb} (see Tables \ref{ftab1}
and \ref{ftab2}). In fit $\fitE$ all bins around the cusp point are
included and $f_{atom}$ is a free parameter, while in fit $\fitF$ seven consecutive
bins centred at $\mm2 = (2m_+)^2$ are excluded and $f_{atom}$ is
fixed to the value given by fit $\fitE$. A comparison of fit $\fitE$ or $\fitF$ with
radiative corrections taken into account (Table \ref{femtab1}) with the
corresponding fits without radiative corrections (fits $\fitE$, $\fitF$ of    
Table \ref{ftab1}) shows that radiative corrections reduce $(a_0-a_2)m_+$
by $\sim9\%$. However, the change in the best fit value of $a_2m_+$ is 
much larger, possibly suggesting again that the determination of this
scattering length is affected by large theoretical uncertainties.

Fits $\fitG$ and $\fitH$ in Tables \ref{femtab1} and \ref{femtab2} are similar to
 $\fitE$ and $\fitF$, respectively, but the constraint between $a_2$ and $a_0$
predicted by analyticity and chiral symmetry \cite{Colangelo:2001sp}
(see Eq. (\ref{ChPT})) is used. A comparison of fits $\fitG$ and $\fitH$ with the
corresponding fits obtained without radiative corrections (fits $\fitG$, $\fitH$ of
Table \ref{ftab1}) shows that radiative corrections reduce $a_0m_+$ by
$\sim6\%$.

For all fits $\fitG$ to $\fitH$ in Tables \ref{femtab1} and \ref{femtab2} the
effect of changing the maximum allowed photon energy $E_{cut}$ from
0.005 to 0.020~\geve ~is found to be negligible.

\begin{table*}
\caption{Fit results with electromagnetic corrections: $\pi \pi$ scattering parameters.
Parameter values without errors have been kept fixed in the fit or calculated using the constraint between $a_2$ and $a_0$ given by Eq. (\ref{ChPT}).}
\begin{center} 
\begin{tabular}{|l|l|l|l|l|l|} \hline
Fit & $\chi^2/NDF$ & $a_0 m_+$ & $a_2 m_+$ & $(a_0-a_2) m_+$ & $f_{atom}$ \\ \hline
$ \fitA $ &  205.6/195 &  $  0.2391$(56) & $ -0.0092$(91) & $  0.2483$(45) & $  0.0625$(92) \\
$ \fitB $ &  202.9/189 &  $  0.2400$(59) & $ -0.0061$(98) & $  0.2461$(49) & $  0.0625$ \\
$ \fitC $ &  222.1/196 &  $  0.2203$(28) & $ -0.0443$ & $  0.2646$(21) & $  0.0420$(77) \\
$ \fitD $ &  219.7/190 &  $  0.2202$(28) & $ -0.0444$ & $  0.2645$(22) & $  0.0420$ \\
$ \fitE $ &  477.4/452 &  $  0.2330$(92) & $ -0.0241$(129) & $  0.2571$(48) & $  0.0631$(97) \\
$ \fitF $ &  474.4/446 &  $  0.2350$(97) & $ -0.0194$(140) & $  0.2544$(53) & $  0.0631$ \\
$ \fitG $ &  479.8/453 &  $  0.2186$(32) & $ -0.0447$ & $  0.2633$(24) & $  0.0538$(77) \\
$ \fitH $ &  478.1/447 &  $  0.2178$(33) & $ -0.0449$ & $  0.2627$(25) & $  0.0538$ \\
\hline \end{tabular}
\end{center}

\label{femtab1}
\end{table*}

\begin{table*}
\caption{Fit results with electromagnetic corrections: coefficients of the tree-level $K \rightarrow 3\pi$  weak decay amplitudes.
Parameter values without errors have been kept fixed in the fit.}
\begin{center} 
\begin{tabular}{|l|l|l|l|l|l|l|} \hline
Fit & $g_0$ & $h_0$ & $k_0$ & $g$ & $h$ & $k$ \\ \hline
$ \fitA $ &  $  0.6453$(22) & $ -0.0355$(18) & $  0.0099$ & $ -0.2112$ & $  0.0067$ & $ -0.0048$ \\
$ \fitB $ &  $  0.6446$(23) & $ -0.0352$(18) & $  0.0099$ & $ -0.2112$ & $  0.0067$ & $ -0.0048$ \\
$ \fitC $ &  $  0.6525$(9) & $ -0.0433$(8) & $  0.0099$ & $ -0.2112$ & $  0.0067$ & $ -0.0048$ \\
$ \fitD $ &  $  0.6526$(9) & $ -0.0432$(8) & $  0.0099$ & $ -0.2112$ & $  0.0067$ & $ -0.0048$ \\
$ \fitE $ &  $  0.6293$(47) & $ -0.0445$(46) & $  0.0085$ & $ -0.1928$(23) & $ -0.0000$(20) & $ -0.0090$(20) \\
$ \fitF $ &  $  0.6311$(51) & $ -0.0429$(49) & $  0.0085$ & $ -0.1938$(25) & $  0.0004$(20) & $ -0.0089$(20) \\
$ \fitG $ &  $  0.6219$(9) & $ -0.0520$(9) & $  0.0085$ & $ -0.1894$(4) & $ -0.0003$(20) & $ -0.0077$(19) \\
$ \fitH $ &  $  0.6220$(9) & $ -0.0521$(9) & $  0.0085$ & $ -0.1895$(4) & $ -0.0002$(20) & $ -0.0077$(19) \\
\hline \end{tabular}
\end{center}

\label{femtab2}
\end{table*}

\begin{sloppypar}

No study of radiative corrections has been performed in the framework
of the CI approach \cite{Cabibbo:2005ez}. 
However, the dominating radiative effects (Coulomb interaction and 
photon emission) are independent of the specific approximation.
Therefore, extracting the relative effect of radiative corrections from 
the BB calculation and using it for the fit to the CI formula 
is justified. In order to obtain an approximate estimate of radiative effects in 
this case, we have corrected the fit procedure by multiplying
the absolute value of the $\kcnn$ decay amplitude given in
ref. \cite{Cabibbo:2005ez} by $|A_{00+}^{rad}/A_{00+}|$ \cite{Isidori:2009priv}, as obtained
in the framework of the BB formulation \cite{Colangelo:2006va,Bissegger:2008ff}. 
Because of the non-physical singularity
of $A_{00+}^{rad}$ at $\mm2 = (2m_+)^2$ in the BB formulation, in
the calculation of the $\kcnn$ decay amplitude for the $51^{st}$ bin
we also multiply the squared amplitude of ref. \cite{Cabibbo:2005ez} 
by $1+f_{atom}$.
\end{sloppypar}

The results of these radiative-corrected fits to the $\mm2$ distribution from $\kcnn$ decay
performed using the CI formula are listed in Tables \ref{femtab1} and  
\ref{femtab2} (Fits $\fitA$ to $\fitD$).
The parameter correlations for two fits which include electromagnetic effects 
are shown in Tables \ref{cabicorr_A} and \ref{berncorr_E}. 

\begin{table}
\caption{Fit parameter correlations for the CI formulation with radiative correction (fit $\fitA$ in Table \ref{femtab1}).}
\begin{tabular}{|l|rrrrr|} \hline
       & $g_0$ & $h_0$ & $a_0-a_2$ & $a_2$ & $f_{atom}$  \\ \hline 
$g_0$  & $ 1.000$ &  &  &  &   \\
$h_0$  & $-0.629$ & $ 1.000$ &  &  &   \\
$a_0-a_2$  & $ 0.794$ & $-0.719$ & $ 1.000$ &  &   \\
$a_2$  & $-0.913$ & $ 0.883$ & $-0.873$ & $ 1.000$ &   \\
$f_{atom}$  & $-0.516$ & $ 0.387$ & $-0.650$ & $ 0.542$ & $ 1.000$  \\
\hline 
\end{tabular}

\label{cabicorr_A}
\end{table}

\begin{table}
\caption{Fit parameter correlations for the BB formulation with radiative correction (fit $\fitE$ in Table \ref{femtab1}).}
\resizebox{\columnwidth}{!}{%
\begin{tabular}{|l|rrrrrrrr|} \hline
       & $g_0$ & $h_0$ & $ g$ & $ h$ & $ k$ & $f_{atom}$ & $a_0-a_2$ & $a_2$  \\ \hline 
$g_0$  & $ 1.000$ &  &  &  &  &  &  &   \\
$h_0$  & $ 0.997$ & $ 1.000$ &  &  &  &  &  &   \\
$ g$  & $-0.972$ & $-0.965$ & $ 1.000$ &  &  &  &  &   \\
$ h$  & $ 0.234$ & $ 0.220$ & $-0.255$ & $ 1.000$ &  &  &  &   \\
$ k$  & $-0.211$ & $-0.225$ & $ 0.194$ & $ 0.889$ & $ 1.000$ &  &  &   \\
$f_{atom}$  & $ 0.597$ & $ 0.570$ & $-0.652$ & $ 0.172$ & $-0.111$ & $ 1.000$ &  &   \\
$a_0-a_2$  & $-0.870$ & $-0.843$ & $ 0.934$ & $-0.404$ & $-0.001$ & $-0.682$ & $ 1.000$ &   \\
$a_2$  & $ 0.977$ & $ 0.982$ & $-0.976$ & $ 0.141$ & $-0.310$ & $ 0.597$ & $-0.839$ & $ 1.000$  \\
\hline 
\end{tabular}

}
\label{berncorr_E}
\end{table}

\begin{figure}
\begin{center}
\resizebox{1.05\columnwidth}{!}{%
\setlength{\unitlength}{1mm}
\begin{picture}(120.,120.)
\includegraphics[width=120mm]{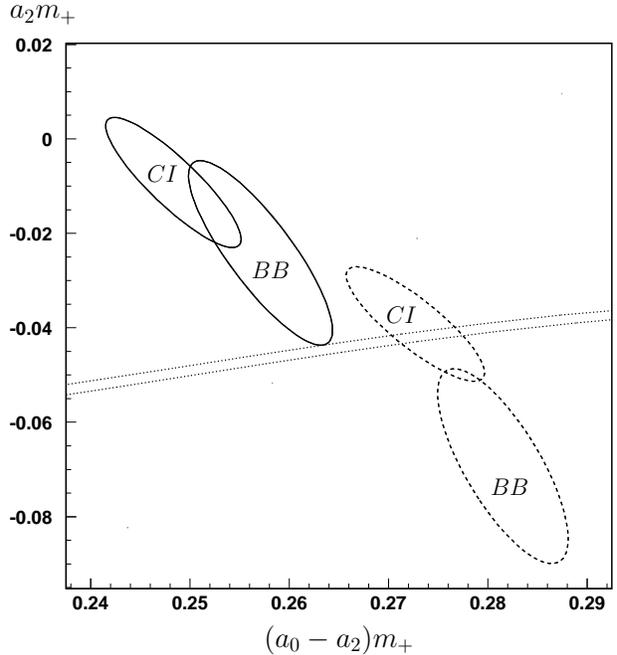}
\put(-112.,113.){\makebox(0,0){\Large $a_2 m_+ $}}
\put(-60.,3.){\makebox(0,0){\Large $(a_0-a_2) m_+ $}}

\put(-49.,60.3){\makebox(0,0){\large $\fitA$}}
\put(-91.,85.){\makebox(0,0){\large $\fitA$}}
\put(-72.,68.){\makebox(0,0){\large $\fitE$}}
\put(-30.,30.){\makebox(0,0){\large $\fitE$}}

\end{picture}
}
\end{center}
\caption{
$68\%$ confidence level ellipses taking into account the statistical uncertainties only. 
Dashed line  ellipses:  fits $\fitA$ and $\fitE$ without radiative corrections.
Solid line ellipses: fits $\fitA$ and $\fitE$ with radiative corrections.
The theoretical band allowed by the ChPT constraint (see Eq. (\ref{ChPT})) 
is shown by the dotted curves.
}

\label{cuspa02}
\end{figure}

Fig. \ref{cuspa02} illustrates the fit results for the fits $\fitA$ and $\fitE$ with 
and without radiative corrections. All the fits are performed using the same $\kcnn$ 
data sample.

\subsection{Pionium formation and other electromagnetic effects at the cusp point}
Pionium formation in particle decay and in charged particle scattering was 
studied in early theoretical work \cite{Silagadze:1994wr,Wycech:1993ci},
but a unified description of its production together with other 
electromagnetic effects near threshold was missing.

\begin{sloppypar}
In a more recent approach \cite{Gevorkyan:2006rh}, electromagnetic effects in
$\kcnn$ decay have been studied in the framework of nonrelativistic quantum
mechanics using a potential model to describe the electromagnetic interaction
between the $\pi^+ \pi^-$ pair in loop diagrams. This model is equivalent to
a perturbative one, in which all simple sequential $\pi^+ \pi^-$ loops with
electromagnetic interactions between the two charged pions are taken into 
account to all orders (including the formation of electromagnetically bound 
final states), but there is no emission of real photons and the 
electromagnetic interaction with the other $\pi^{\pm}$ from the $\kccc$ 
decay is ignored. Because  of these limitations, the model of 
ref. \cite{Gevorkyan:2006rh} cannot be directly applied to the full physical 
region of the $\kcnn$ decay; however, contrary to the BB formulation 
\cite{Bissegger:2008ff}, its integral effect over a narrow region which 
includes the cusp point ($\mm2=4m_+^2$) can be calculated.
\end{sloppypar}

We have implemented the electromagnetic effects predicted by the model of ref.
\cite{Gevorkyan:2006rh} in the parameterization of the CI formulation 
\cite{Cabibbo:2005ez} (the detailed procedure is described in Eqs. (6, 7, 8) 
of ref. \cite{Gevorkyan:2007ki}). In the theoretical $\mm2$ distribution
the electromagnetic correction for the bin centred at $4m_+^2$ (bin 51),
averaged over the bin, depends on the bin width, as it includes contributions
from both pionium bound states with negligible widths and a very narrow 
peak of unbound $\pi^+ \pi^-$ states annihilating to $\pi^0 \pi^0$. For
the bin width of 0.00015~\gevmsq ~used in the fits, these effects
increase the content of bin 51 by 5.8$\%$, in agreement with the results 
of the fits performed using $f_{atom}$ as a free parameter (see Tables 
\ref{ftab1}, \ref{femtab1}). Thus the model of ref. \cite{Gevorkyan:2006rh}
explains why the typical fit result for $f_{atom}$ is nearly twice as large as 
the prediction for pionium contribution only, as calculated in refs. 
\cite{Pilkuhn:1978rf,Silagadze:1994wr}.

Near the cusp point the two calculations of electromagnetic effects
\cite{Bissegger:2008ff} and  \cite{Gevorkyan:2006rh,Gevorkyan:2007ki} are very similar 
numerically, thus increasing the confidence in the central cusp bin radiative effect 
calculated using Eq. (8) of ref. \cite{Gevorkyan:2007ki}. However, at larger
distances from the cusp the approach of refs. 
\cite{Gevorkyan:2006rh,Gevorkyan:2007ki} leads to deviations from the
electromagnetic corrections of ref. \cite{Bissegger:2008ff}. This can be 
explained by the fact that the model of ref. \cite{Gevorkyan:2006rh} 
takes into account only processes that dominate near the cusp point. 
For this reason we do not use 
this model in the fits, but we consider it as a complementary calculation 
limited to a region very close to the cusp point, providing a
finite result for the bin centred at $\mm2 = 4m_+^2$ which 
the formulation of ref. \cite{Bissegger:2008ff} does not provide.

\section{Systematic uncertainties}
\label{systematics}

As shown below, all systematic corrections affecting the best fit values of
the coefficients describing the $\kcnn$ weak amplitude at tree level,
$g_0$ and $h_0$ (see Eq.~(\ref{amp0})), are found to be much smaller
than the statistical errors. We use these corrections as additional
contributions to the systematic uncertainties instead of correcting
the central values of these parameters.

For a given fit, we find that the systematic uncertainties affecting the best 
fit parameters do not change appreciably if the fit is performed with or
without electromagnetic corrections. In addition, we find that, with
the exception of $f_{atom}$, the systematic uncertainties affecting all other
parameters are practically the same if in the fit the seven consecutive 
bins centred at $\mm2=4m_+^2$ are included (and $f_{atom}$ is used as a free 
parameter), or if they are excluded (and the value of $f_{atom}$ is fixed).

For these reasons, we give detailed estimates of the systematic uncertainties
only for fits $\fitA$, $\fitC$, $\fitE$, $\fitG$ performed with the decay amplitude corrected 
for electromagnetic effects. 

\begin{sloppypar}
The parameters $g,h,k$ which describe the $\kccc$ weak amplitude at
tree level are used as free parameters when fitting the data to the  
BB formulation \cite{Colangelo:2006va,Bissegger:2008ff}. However, they
enter into the $\kcnn$ decay amplitude only through rescattering terms, 
thus we do not consider the best fit values of these parameters as a 
measurement of physically important values. Here we do not estimate the 
systematic uncertainties affecting them and we discuss the
uncertainties associated with $\kccc$ decay in Section \ref{external}.
In the study of the systematic uncertainties affecting the $\kcnn$ decay 
parameters we fix the values of the $\kccc$ decay parameters $g,h,k$ in
the BB formulation to their best fit values shown in Table \ref{femtab2}.
\end{sloppypar}

The fit interval for the presentation of the final results (bins 26--226 of width 
0.00015 \gevmsq, with bin 51 centred at $4m_{\pi^+}^2$) has been 
chosen to minimize the total experimental error of the measured $a_0-a_2$. 
If the upper limit of the fit region, $s_3^{max}$, is increased, the 
statistical error decreases. All our fits give good $\chi^2$  up to rather 
high $s_3^{max}$ values where the acceptance is small \footnote{At the maximum 
kinematically allowed $s_3$ value the $\pi^{\pm}$ is at rest in the $K^{\pm}$ 
decay frame. In this case, it moves along the $K^{\pm}$ flight path inside the 
beam vacuum tube and cannot be detected. Near this maximum $s_3$ 
value the acceptance is very sensitive to the precise beam shape and position 
due to the $\pi^{\pm}$ narrow angular distribution, and it is difficult to 
reproduce it in the Monte-Carlo simulation.}. However, the systematic error 
increases with $s_3^{max}$, especially the contributions from trigger 
inefficiency and non-linearity of the LKr response. The total experimental 
error on $a_0-a_2$, obtained by adding quadratically the statistical and 
systematic error, has a minimum when the upper limit of the fit interval 
corresponds to bin 226.

\subsection{Acceptance}

The detector acceptance to $\kcnn$ decays depends strongly on the position of 
the $K^{\pm}$ decay vertex along the nominal beam axis, $Z$, so the $Z$ 
distribution provides a sensitive tool to control the quality 
of the acceptance simulation.

\begin{figure}[ht]
\begin{center}
\resizebox{1.05\columnwidth}{!}{%
\setlength{\unitlength}{1mm}
\begin{picture}(100.,100.)

\includegraphics[width=100mm]{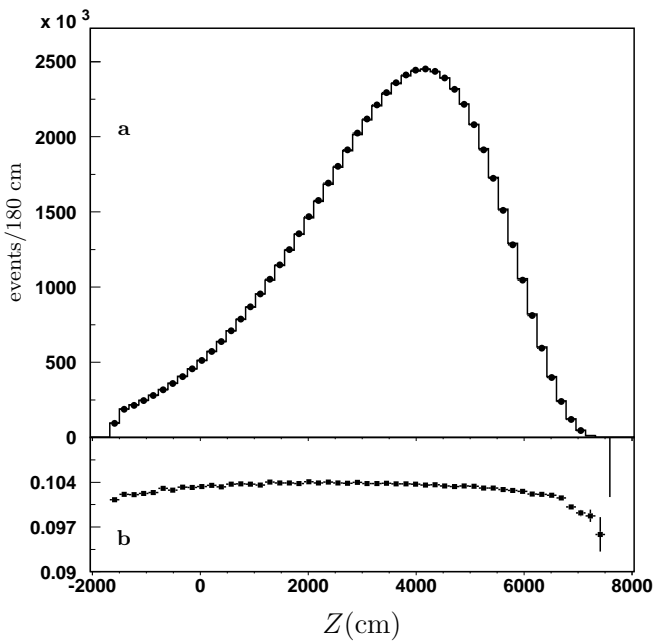}
\put(-50.,2.){\makebox(0,0){\large $Z$(cm)}}
\put(-85.,75.){\makebox(0,0){\bf a}}
\put(-85.,15.){\makebox(0,0){\bf b}}
\put(-101.,60.){\makebox(0,0){\rotatebox{90}{events/180 cm}}}
\end{picture}
}
\end{center}

\caption{
$\kcnn$ decay $Z$ distributions for data and Monte-Carlo simulation. 
{\bf a}: Experimental (solid circles) and simulated (histogram) 
distributions, normalized to experimental statistics. 
{\bf b}: Ratio between the experimental and simulated distributions.
The nominal position of LKr front face is at $Z=12108.2$~cm.    
}
\label{zdist}
\end{figure}

Fig. \ref{zdist} shows the comparison between the data and Monte-Carlo
simulated $Z$ distributions. The small difference between the shapes of
the two distributions in the region $Z<0$ disappears when the trigger 
efficiency correction is applied, so this difference is taken into account
in the contribution to the systematic uncertainties from the trigger
efficiency (see Tables \statsys). 

A small difference between the shapes of the two distributions is also present
in the large $Z$ region in the area where the acceptance drops 
because of the increasing probability for the charged pion track 
to cross the spectrometer too close to the event COG. The effect of this
acceptance difference has been checked by introducing a small mismatch
in the track radius cuts between real and simulated data, and also 
by applying small changes to the  LKr energy scale (equivalent to shifts
of the event $Z$ position similar to the effect observed in the acceptance). 
The corresponding small changes of the fit results are considered as the 
acceptance related contribution to the systematic uncertainties (quoted as
Acceptance(Z) in Tables \statsys).     

The Monte Carlo sample from which the acceptance and resolution effects used
in the fits are derived, is generated under the assumption that the $\kcnn$ 
matrix element, ${\cal M}$, depends only on $u$. We have studied the sensitivity
of the fit results to the presence of a $v$-dependent term by adding  
to $|{\cal M}|^2$ a term of the form $k_0 v^2$ or $k^\prime Re({\cal M})v^2$, 
consistent with the observed $v$ dependence in the data. The largest 
variations of the fit results are shown in Tables \statsys \ as the 
contributions to the systematic uncertainties arising from the 
simplified matrix element used in the Monte Carlo (they are quoted as
Acceptance(V)).

\subsection{Trigger efficiency}

During data taking in 2003 and 2004 some changes to the 
trigger conditions were introduced following improvements in detector and
electronics performance. In addition, different minimum bias triggers with 
different downscaling factors were used. As a consequence, trigger effects 
have been studied separately for the data samples taken during seven periods
of uniform trigger conditions. Details of the trigger efficiency for the 
$\kcnn$ decay events are given in \cite{Batley:2006tt,Batley:2007yfa}.

As described in Section \ref{evselrec}, $\kcnn$ events were recorded by
a first level trigger using signals from the scintillator hodoscope
(Q1) and LKr (NUT), followed by a second level trigger using drift
chamber information (MBX). Events were also recorded using other
triggers with different downscaling factors for different periods: a minimum bias
NUT trigger (ignoring both Q1 and MBX); and a minimum bias Q1*MBX trigger
(ignoring LKr information). Using the event samples recorded with these downscaled 
triggers, and selecting $\kcnn$ decays as described in section
\ref{evselrec}, it was possible to measure separately two efficiencies: 

\begin{enumerate}
\item the efficiency of the minimum bias Q1*MBX trigger using the event sample 
recorded by the minimum bias NUT trigger;
\item the efficiency of the minimum bias NUT trigger using the events recorded by 
the minimum bias Q1*MBX trigger. 
\end{enumerate}

These two efficiencies were multiplied together to obtain the full trigger 
efficiency.

\begin{figure}[ht]
\begin{center}
\resizebox{1.05\columnwidth}{!}{%
\setlength{\unitlength}{1mm}
\begin{picture}(120.,120.)
\includegraphics[width=120mm]{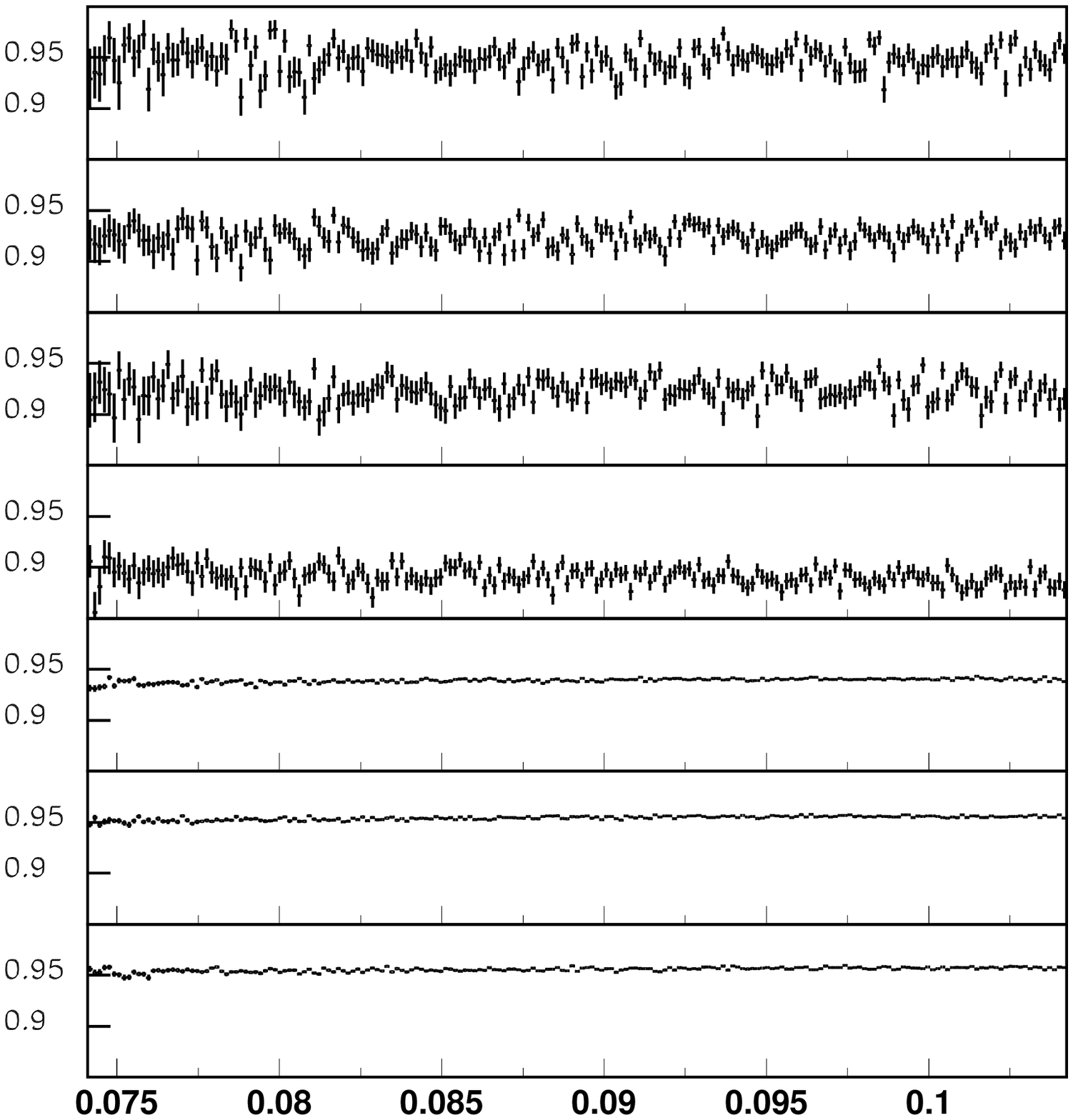}
\put(-8.,110.){\makebox(0,0){\bf a}}
\put(-8.,95.){\makebox(0,0){\bf b}}
\put(-8.,80.){\makebox(0,0){\bf c}}
\put(-8.,65.){\makebox(0,0){\bf d}}
\put(-8.,50.){\makebox(0,0){\bf e}}
\put(-8.,35.){\makebox(0,0){\bf f}}
\put(-8.,20.){\makebox(0,0){\bf g}}
\put(-60.,3.){\makebox(0,0){\large $\mm2$ \gevmsq}}
\put(-121.,70.){\makebox(0,0){\bf \Large $\epsilon$}}
\end{picture}
}
\end{center}
\caption{
Trigger efficiency $\epsilon$ as a function of $\mm2$ for the different time periods with different trigger
conditions ({\bf a}--{\bf c}: 2003, {\bf d}--{\bf g}: 2004). The errors are defined by the available statistics of the
event samples recorded by the two minimum bias triggers. 
}
\label{trigeff}
\end{figure}

The measured efficiencies for seven different periods are shown in Fig.~\ref{trigeff}
as a function of the reconstructed $\mm2$. In the initial data taking periods 
the samples of minimum bias events were rather small, resulting in relatively 
large statistical errors. However, we can improve the estimate of the trigger 
efficiency for these periods under the additional assumption that it
is a smooth function of $\mm2$ (this assumption is justified by the fact that no  
anomaly is expected nor observed in its behaviour). We find that a 2-nd degree 
polynomial
\begin{equation}
p_0 + p_1*(\mm2-4m_+^2) + p_2*(\mm2-4m_+^2)^2
\label{trigap}
\end{equation}
describes well the trigger efficiency over the $\mm2$ fit interval. Moreover,
over this interval the dependence is almost linear, so we expect a
negligible effect on the determination of the scattering lengths.

Fits are made separately for each of the data taking periods shown in
Fig. \ref{trigeff}. In a first fit, the $\mm2$ distribution from the
data and the corresponding trigger efficiency are fitted simultaneously,  
and the theoretical $\mm2$ distribution, distorted 
by the acceptance and resolution effects, is multiplied by the corresponding
trigger efficiency, as parameterized using Eq.~(\ref{trigap}). The fit to
the $\mm2$ distribution alone is then repeated under the assumption of a fully efficient 
trigger, and the results of the two fits are compared to obtain the trigger
efficiency correction and its effective error. As an example, Table~\ref{trigtable} 
lists the trigger corrections to the best fit parameters of fits $\fitA$ and $\fitC$ 
(see Table~\ref{femtab1}).

\begin{table}
\caption{Trigger efficiency corrections for the best fit parameters of
fits  $\fitA$ and $\fitC$ of Table~\ref{femtab1}.}
\begin{center}
\begin{tabular}{|l|r|r|}                      \hline
                              &  fit $\fitA$                                 &  fit   $\fitC$                  \\ \hline
$g_0$                    &  $0.00056(81)$                 & $0.00111(70)$     \\ \hline
$h_0$                    &  $0.00136(95)$                 & $0.00136(66)$     \\ \hline
$(a_0-a_2) m_+$   &  $-0.00041(67)$                &             -               \\ \hline
$a_0 m_+$             &            -                             & $0.00065(48)$      \\ \hline
$a_2 m_+$             &  $0.00226(190)$               &               -              \\ \hline
$f_{atom}$             &  $0.00070(86)$                 & $-0.00049(82)$    \\ \hline
\end{tabular}          
\end{center}
\label{trigtable}
\end{table}

The trigger corrections are all in agreement with  
zero within their statistical uncertainties.
For a conservative estimate, we combine in quadrature  
the corrections and their errors to
obtain the trigger efficiency contribution to the systematic
uncertainties of the best fit results (see Tables \statsys). 

\subsection{LKr resolution}

As described in Section \ref{evselrec}, the $\pi^0 \pi^0$ invariant
mass $M_{00}$ is determined using only information from the LKr
calorimeter (photon energies and coordinates of their impact points). 
 The measurement of the scattering lengths relies, therefore, on the correct 
description of the $M_{00}$ resolution in the Monte Carlo simulation. 

\begin{figure}[ht]
\begin{center}
\resizebox{1.05\columnwidth}{!}{%
\setlength{\unitlength}{1mm}
\begin{picture}(100.,100.)

\includegraphics[width=100mm]{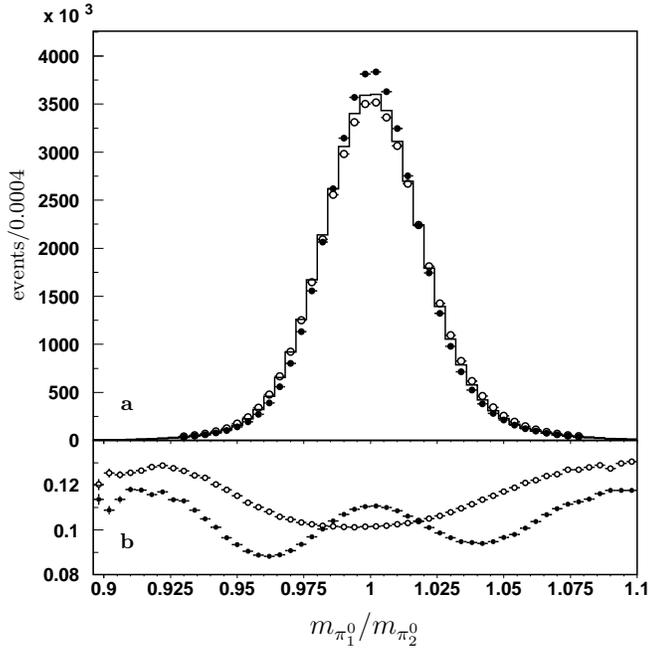}
\put(-50.,2.){\makebox(0,0){\large $m_{\pi^0_1}/m_{\pi^0_2}$}}
\put(-85.,35.){\makebox(0,0){\bf a}}
\put(-85.,15.){\makebox(0,0){\bf b}}
\put(-101.,60.){\makebox(0,0){\rotatebox{90}{events/0.0004}}}
\end{picture}
}
\end{center}

\caption{
Distributions of the measured ratio $m_{\pi^0_1}/m_{\pi^0_2}$ (see text) for 
the data of 2004. {\bf a}:~solid circles - data events; open circles -  data events with
the LKr cluster energies artificially smeared as described in the text; 
histogram - simulated distribution, normalized to data statistics. 
{\bf b}:~corresponding ratios of data and simulated distributions.
}
\label{resol}
\end{figure}

In order to check the quality of the LKr energy resolution we cannot use the $\pi^0$ mass 
peak in the two-photon invariant mass distribution, because the
nominal $\pi^0$ mass \cite{Amsler:2008zz} is used in the
reconstruction of the two-photon decay vertex (see Section \ref{evselrec}).
We find that a convenient variable which is sensitive to all random fluctuations of the LKr
response, and hence to its energy resolution, is the ratio $m_{\pi^0_1}/m_{\pi^0_2}$, 
where $m_{\pi^0_1}$ and $m_{\pi^0_2}$ are the measured two-photon
invariant masses for the more and less energetic $\pi^0$, respectively,
in the same $\kcnn$ decay. The distributions of this ratio for real and simulated 
events are shown in Fig. \ref{resol}. One can see that the width of
the distribution for simulated events is slightly larger than that of
the data: the rms value of the simulated distribution is 0.0216, while it is 0.0211 
for the data. 

In order to check the sensitivity of the fit results to  
a resolution mismatch of this size, we have smeared the measured
photon energies in the data by adding a random energy with a Gaussian distribution
centred at zero and with $\sigma = 0.06$~\geve ~(see Fig. \ref{resol}). Such
a change increases the rms value of the $m_{\pi^0_1}/m_{\pi^0_2}$
distribution from 0.0211 to 0.0224. A fit is then performed for the data 
sample so modified, and the values of the fit parameters are compared with
those obtained using no energy smearing. 

The artificial smearing of the photon energies described above introduces random  
shifts of the fit parameters within their statistical errors. In order
to determine these shifts more precisely than allowed by the
statistics of a single fit, we have repeated the fit eleven times
using for each fit a data sample obtained by smearing the original photon
energies with a different series of random numbers, as described in the
previous paragraph. 
The shifts of the fit parameters, averaged over the eleven fits, represent the
systematic effects, while the errors on those average values are
the corresponding uncertainties. Conservatively, the quadratic
sum of the shifts and their errors is quoted as
``LKr resolution'' in Tables \statsys.

\subsection{LKr non-linearity}

In order to study possible non-linearity effects of the LKr calorimeter response
to low energy photons, we select $\pi^0$ pairs from $\kcnn$ events
using the following criteria:

\begin{enumerate}
\item both $\pi^0 \rightarrow \gamma \gamma$ decays must be close to symmetrical 
($0.45 < \frac{E_{\gamma}}{E_{\pi^0}} < 0.55$);
\item the more energetic $\pi^0$ (denoted as $\pi^0_1$) must fulfil the requirement \\ 
$22$~\geve~$ < E_{\pi^0_1} < 26$~\geve.
\end{enumerate}

\begin{figure}[ht]
\begin{center}
\resizebox{1.05\columnwidth}{!}{%
\setlength{\unitlength}{1mm}
\begin{picture}(100.,100.)

\includegraphics[width=100mm]{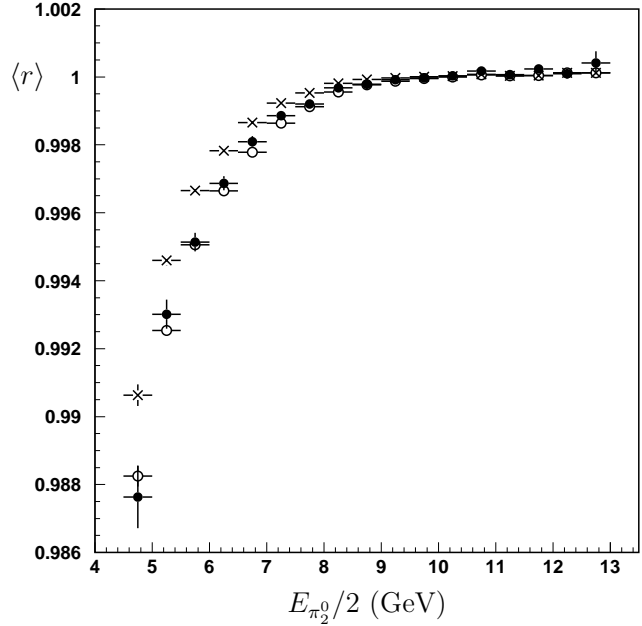}
\put(-50.,2.){\makebox(0,0){\large $E_{\pi^0_2}/2$ (\geve)}}
\put(-100.,80.){\makebox(0,0){\large $\langle r \rangle$}}
\end{picture}
}
\end{center}

\caption{
Average $r = m_{\pi^0_2}/m_{\pi^0_1}$ versus $E_{\pi^0_2}/2$
for $\pi^0$ pairs from $\kcnn$ decays selected as described in the text.  
Solid circles: data; crosses: simulated events; open circles: simulated events
corrected for non-linearity (see text). The $\pi^0_2$ energy is
divided by 2 to compare with the $\gamma$ energy for symmetric $\pi^0$ decays.
}
\label{nonlin}
\end{figure}

For the $\pi^0$ pairs selected in such way we define the ratio of the
two-photon invariant masses, $r=m_{\pi^0_2}/m_{\pi^0_1}$, where $\pi^0_2$
is the lower energy $\pi^0$. Fig. \ref{nonlin} shows the average ratio $\langle r \rangle$ 
as a function of $E_{\pi^0_2}/2$ for both data and simulated events
(for symmetric $\pi^0 \rightarrow \gamma \gamma$ decays
$E_{\pi^0_2}/2$ is the photon energy).    

Because of the resolution effects discussed in the previous
subsection\footnote{The small resolution mismatch between data and simulated
events introduces a negligible effect here.}, 
$\langle r \rangle$ depends on the lowest pion energy even in the case of perfect LKr linearity.
However, as shown in Fig. \ref{nonlin}, for $E_{\pi^0_2}/2 \lesssim 9$~\geve 
~the values of $\langle r \rangle$ for simulated events are systematically above
those of the data, providing evidence for the presence of non-linearity   
effects of the LKr response at low energies.

To study the importance of these effects, we modify all simulated events to 
account for the observed non-linearity multiplying each photon
energy by the ratio $\displaystyle\frac{\langle r_{Data} \rangle}{\langle r_{MC} \rangle}$, where $\langle r_{Data} \rangle$
and $\langle r_{MC} \rangle$ are the average ratios for data and simulated events, respectively. 
As shown in Fig. \ref{nonlin}, the values of $\langle r \rangle$ for the sample of simulated
events so modified are very close to those of the data. The small shifts  
of the best fit parameters obtained using these non-linearity corrections 
are taken as contributions to the systematic uncertainties in Tables \statsys, 
where they are quoted as ``LKr non-linearity''.

\subsection{Hadronic showers in LKR}

The $\pi^{\pm}$ interaction in the LKr may produce multiple energy
clusters which are located, in general, near the impact point of the 
$\pi^{\pm}$ track and in some cases may be identified as photons. To reject 
such ``fake'' photons a cut on the distance $d$ between each photon and
the impact point of any charged particle track at the LKr front face
is implemented in the event selection, as described in Section \ref{evselrec}.
In order to study the effect of these ``fake'' photons on the best fit
parameters we have repeated the fits by varying the cut on the
distance $d$ between 10 and 25~cm in the selection of both data and
simulated $\kcnn$ events. The largest deviations from the results
obtained with the default cut value ($d$=15~cm) are taken as contributions 
to the systematic uncertainties (see Tables \statsys).

\subsection{Other sources}
    
\begin{figure}[ht]
\begin{center}
\resizebox{1.05\columnwidth}{!}{%
\setlength{\unitlength}{1mm}
\begin{picture}(100.,100.)

\includegraphics[width=100mm]{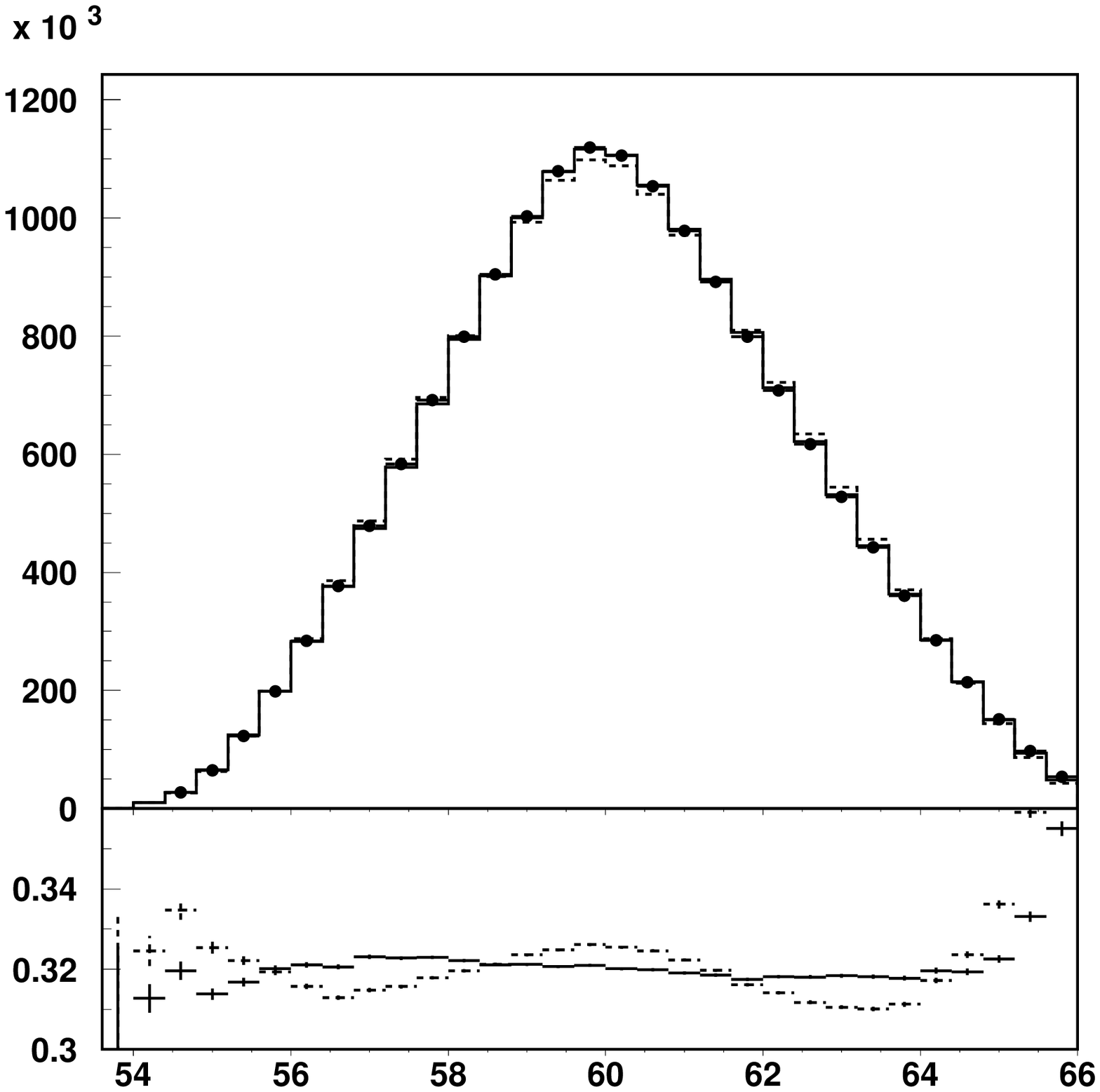}
\put(-50.,2.){\makebox(0,0){\large $P_K$(\gevp)}}
\put(-80.,55.){\makebox(0,0){\bf a}}
\put(-80.,25.){\makebox(0,0){\bf b}}
\put(-101.,60.){\makebox(0,0){\rotatebox{90}{events/0.4 \gevp}}}

\end{picture}
}
\end{center}

\caption{
Distributions of the reconstructed $K^{\pm}$ momentum $P_K$ from the data
and from Monte-Carlo simulation (2003 data). 
{\bf a}: solid circles -- experimental data; dashed line histogram -- simulation; solid line histogram -- simulation with the 
corrected $K^{\pm}$ spectrum width. {\bf b}: corresponding ratios of data and simulated spectra. 
}

\label{pspect}
\end{figure}

The Monte Carlo program includes a complete simulation of the beam magnet
system and collimators with the purpose of reproducing the correlation between
the incident $K^{\pm}$ momenta and trajectories. However, the absolute beam
momentum scale cannot be modelled with the required precision, hence
we tune the average value to the measured ones for each continuous
data taking period (``run'') using $\kccc$ events which are recorded    
during data taking, and also simulated by the Monte Carlo program.

After this adjustment, a residual systematic difference still exists 
between the measured and simulated $K^{\pm}$ momentum distributions,
as shown in Fig. \ref{pspect}. In order to study the sensitivity of the 
best fit parameters to this distribution, we have corrected the width of the 
simulated $K^{\pm}$ momentum distribution to reproduce the measured
distribution (see Fig. \ref{pspect}) using a method based on the rejection of 
simulated events. To minimize the random effect of this rejection, a fraction 
of events has also been removed from the uncorrected MC sample in such a way
that the corrected and uncorrected MC samples have a maximum overlap
of events and the same statistics. The corresponding changes of the
best fit parameters are included in the contributions to the
systematic uncertainties and quoted as ``$P_K$ spectrum'' in Tables \statsys.  

In order to take into account changes of running conditions during
data taking, the number of simulated $\kcnn$ events for each run should be
proportional to the corresponding number of events in the data. However,
because of changes in the trigger efficiency and in acceptance 
related to minor hardware problems, the ratio between the number of 
simulated and real events varies by a few percent during the whole
data taking period. In order to study the effect of the small mismatch
between the two samples on the best fit parameters, we have made them
equal run by run by a random rejection of selected events. The corresponding
shifts of the best fit parameters are considered as a Monte Carlo     
time dependent systematic error, and are listed in Tables \statsys,
where they are quoted as ``MC(T)''.

\begin{table}[ht]
         \caption{Fit parameter systematic uncertainties  in units of $10^{-4}$ for the CI formulation 
                       with electromagnetic corrections (fit $\fitA$ in Table \ref{femtab1}).
                       The factor $m_+$ which should multiply the scattering lengths is omitted for simplicity.}
\begin{center}
\begin{tabular}{|l|r|r|r|r|r|r|} \hline
Source                 & $g_0$    & $h_0$    & $a_0$    & $a_2$    & $a_0-a_2$    & $f_{atom}$  \\ \hline 
Acceptance(Z)          & 22 & 17 & 11 & 14 & 3 & 1 \\ 
Acceptance(V)          & 9 & 3 & 5 & 6 & 1 & 3 \\ 
Trigger efficiency     & 10 & 17 & 22 & 30 & 8 & 11 \\ 
LKr resolution         & 4 & 2 & 11 & 17 & 7 & 56 \\ 
LKr nonlinearity       & 2 & 21 & 39 & 49 & 11 & 5 \\ 
$P_K$ spectrum         & 5 & 3 & 11 & 23 & 12 & 8 \\ 
MC(T)                  & 3 & 2 & 4 & 1 & 5 & 25 \\ 
$k_0$ error            & 8 & 6 & 3 & 4 & 1 & 1 \\ 
Hadronic showers       & 9 & 3 & 3 & 13 & 9 & 20 \\ 
 \hline 
Total systematic       & 29 & 33 & 49 & 67 & 22 & 66 \\ \hline 
Statistical            & 22 & 18 & 56 & 92 & 45 & 93 \\ \hline
\end{tabular}
\end{center}

            \label{cabisi_noskip_atomfree_10}
\end{table}

\begin{table}[ht]
         \caption{Fit parameter systematic uncertainties in units of $10^{-4}$ for the CI formulation 
                       with electromagnetic corrections  and with the ChPT constraint (fit $\fitC$ in Table \ref{femtab1}).
                       The factor $m_+$ which should multiply the scattering lengths is omitted for simplicity.}
\begin{center}
\begin{tabular}{|l|r|r|r|r|r|r|} \hline
Source                 & $g_0$    & $h_0$    & $a_0$    & $a_2$    & $a_0-a_2$    & $f_{atom}$  \\ \hline 
Acceptance(Z)          & 24 & 14 & 4 & 1 & 3 & 9 \\ 
Acceptance(V)          & 8 & 4 & 2 & 0 & 2 & 0 \\ 
Trigger efficiency     & 13 & 15 & 8 & 2 & 6 & 10 \\ 
LKr resolution         & 0 & 2 & 2 & 0 & 1 & 46 \\ 
LKr nonlinearity       & 12 & 13 & 13 & 3 & 10 & 31 \\ 
$P_K$ spectrum         & 0 & 0 & 2 & 1 & 2 & 5 \\ 
MC(T)                  & 2 & 2 & 6 & 1 & 4 & 24 \\ 
$k_0$ error            & 7 & 7 & 1 & 0 & 0 & 2 \\ 
Hadronic showers       & 5 & 3 & 4 & 1 & 3 & 19 \\ 
 \hline 
Total systematic       & 33 & 26 & 18 & 4 & 14 & 65 \\ \hline 
Statistical            & 9 & 8 & 28 & 6 & 21 & 77 \\ \hline
\end{tabular}
\end{center}

            \label{cabisi_chpt_noskip_atomfree_10}
\end{table}

\begin{table}[ht]
         \caption{Fit parameter systematic uncertainties in units of $10^{-4}$ for the BB formulation 
                       with electromagnetic corrections  (fit $\fitE$ in Table \ref{femtab1}).
                       The factor $m_+$ which should multiply the scattering lengths is omitted for simplicity.}
\begin{center}
\begin{tabular}{|l|r|r|r|r|r|r|} \hline
Source                 & $g_0$    & $h_0$    & $a_0$    & $a_2$    & $a_0-a_2$    & $f_{atom}$  \\ \hline 
Acceptance(Z)          & 31 & 21 & 16 & 20 & 4 & 0 \\ 
Acceptance(V)          & 6 & 1 & 7 & 8 & 1 & 4 \\ 
Trigger efficiency     & 26 & 22 & 29 & 39 & 10 & 13 \\ 
LKr resolution         & 10 & 9 & 21 & 29 & 9 & 60 \\ 
LKr nonlinearity       & 34 & 36 & 56 & 67 & 12 & 1 \\ 
$P_K$ spectrum         & 12 & 11 & 18 & 32 & 13 & 10 \\ 
MC(T)                  & 2 & 1 & 4 & 1 & 5 & 25 \\ 
$k_0$ error            & 5 & 5 & 4 & 6 & 2 & 1 \\ 
Hadronic showers       & 2 & 4 & 8 & 18 & 10 & 20 \\ 
 \hline 
Total systematic       & 56 & 50 & 72 & 94 & 25 & 70 \\ \hline 
Statistical            & 47 & 46 & 92 & 129 & 48 & 97 \\ \hline
\end{tabular}
\end{center}

             \label{bernfit_noskip_atomfree_10}
\end{table}

\begin{table}[ht]
         \caption{Fit parameter systematic uncertainties in units of $10^{-4}$ for the BB formulation 
                       with electromagnetic corrections and  with the ChPT constraint (fit $\fitG$ in Table \ref{femtab1}).
                       The factor $m_+$ which should multiply the scattering lengths is omitted for simplicity.}
\begin{center}
\begin{tabular}{|l|r|r|r|r|r|r|} \hline
Source                 & $g_0$    & $h_0$    & $a_0$    & $a_2$    & $a_0-a_2$    & $f_{atom}$  \\ \hline 
Acceptance(Z)          & 24 & 14 & 4 & 1 & 3 & 9 \\ 
Acceptance(V)          & 8 & 4 & 2 & 1 & 2 & 0 \\ 
Trigger efficiency     & 14 & 16 & 9 & 2 & 7 & 8 \\ 
LKr resolution         & 0 & 1 & 2 & 1 & 2 & 46 \\ 
LKr nonlinearity       & 12 & 13 & 13 & 3 & 10 & 31 \\ 
$P_K$ spectrum         & 0 & 0 & 2 & 1 & 2 & 5 \\ 
MC(T)                  & 2 & 2 & 6 & 1 & 4 & 24 \\ 
$k_0$ error            & 7 & 7 & 0 & 0 & 0 & 2 \\ 
Hadronic showers       & 5 & 3 & 4 & 1 & 3 & 17 \\ 
 \hline 
Total systematic       & 33 & 26 & 18 & 4 & 14 & 64 \\ \hline 
Statistical            & 9 & 9 & 32 & 8 & 24 & 77 \\ \hline
\end{tabular}
\end{center}

             \label{bernfit_chpt_noskip_atomfree_10}
\end{table}

\section{External uncertainties}
\label{external}

The most important source of external error is the value of  $|A_+|$, 
obtained from  the measured ratio of the $\kccc$ and $\kcnn$ decay 
rates, $R = 3.175 \pm 0.050$ \cite{Amsler:2008zz}. This ratio is
proportional to $|A_+|^2$, so 
$$\delta |A_+|/|A_+| = 0.5 (\delta R)/R.$$
The typical $|A_+|$ uncertainty is, therefore, $\delta |A_+| \approx 0.015$.

We have checked the shifts of the fit results due to the variation of $|A_+|$ within its uncertainty. 
Each fit is redone twice changing the $|A_+|$ value by $+ \delta |A_+|$ and $- \delta |A_+|$. One half
of the variation of the fit parameters corresponding to these two fits is listed in Table \ref{ext_err},
and is taken as the external contribution to the full parameter uncertainty.   

\begin{table}[ht]
\caption{Contributions to the fit parameter uncertainties (in units of $10^{-4}$) due to the external error $ \delta |A_+|$.}
\begin{center} 

\begin{tabular}{|l|r|r|r|r|r|r|} \hline
Fit  &   $g_0$   & $h_0$ &    $a_0 m_+$  &    $a_2 m_+$      & $(a_0-a_2) m_+$  &    $f_{atom}$ \\ \hline
$\fitA$    &   3   &   0   &   27  &   14    &  13       &    1    \\
$\fitC$    &   1   &   2   &   24  &   6      &   18      &    5    \\
$\fitE$    &   5   &   3   &   32  &   18    &   14      &    1    \\
$\fitG$    &   0   &   2   &   25  &   6      &  19       &    5    \\
\hline 
\end{tabular}

\end{center}
\label{ext_err}
\end{table}

\section{\boldmath $\pi \pi$ \unboldmath scattering lengths: final results}
\label{final}

The BB formulation with radiative corrections \cite{Bissegger:2008ff}
provides presently the most complete description of rescattering
effects in $K \rightarrow 3\pi$ decay. For this reason we use the
results from the fits to this formulation to present our final results on
the $\pi \pi$ scattering lengths:
\begin{eqnarray}
\nonumber
(a_0-a_2)m_+ = 0.2571\pm0.0048(stat.) \\
\pm0.0025(syst.)\pm0.0014(ext.);
\label {a0a2_uncon}
\end{eqnarray}
\begin{eqnarray}
\nonumber
a_2m_+ = -0.024\pm0.013(stat.) \\
\pm0.009(syst.)\pm0.002(ext.).
\label {a2_uncon}
\end{eqnarray} 
The values of the $\pi \pi$ scattering lengths,
$(a_0-a_2)m_+$ and $a_2m_+$, are obtained from fit $\fitE$ of Table
\ref{femtab1}. In addition to the statistical, systematic and external
errors discussed in the previous sections, these values are affected
by a theoretical uncertainty.
We note that, at the level of approximation of the BB and CI amplitude
expression used in the fits, a difference of 0.0088(3.4\%)is found 
between the values of $(a_0-a_2)m_+$ and of 0.015(62\%) for $a_2m_+$.
For the sake of comparison with other independent results on the $\pi \pi$
scattering lengths we take into account these differences as theoretical
uncertainty.

From the measurement of the lifetime of pionium by the DIRAC 
experiment at the CERN PS \cite{Adeva:2005pg} a value of 
$|a_0-a_2|m_+ = 0.264^{+0.033}_{-0.020}$ 
was deduced which agrees, within its quoted uncertainty, with our result
(it should be noted that this measurement provides only a determination 
of $|a_0-a_2|$, while our measurement of $\kcnn$ decay is also sensitive 
to the sign).

\begin{figure}
\begin{center}
\resizebox{1.05\columnwidth}{!}{%
\setlength{\unitlength}{1mm}
\begin{picture}(120.,120.)
\includegraphics[width=120mm]{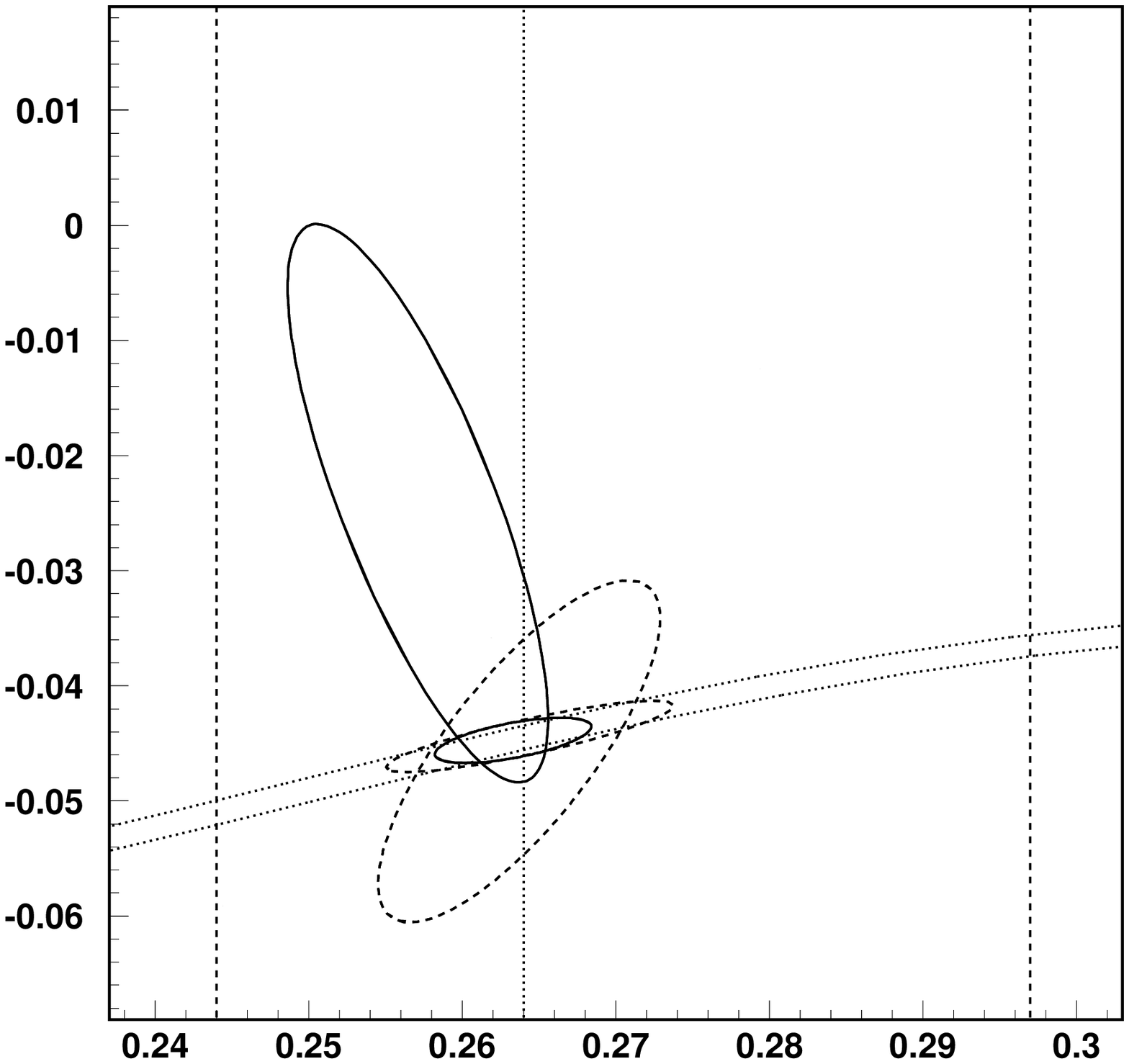}
\put(-112.,113.){\makebox(0,0){\Large $a_2 m_+$}}
\put(-60.,3.){\makebox(0,0){\Large $(a_0-a_2) m_+$}}

\put(-52.5,60.3){\makebox(0,0){NA48/2}}
\put(-52.5,55.5){\makebox(0,0){\large $K_{e4}$}}
\put(-81.,65.){\makebox(0,0){NA48/2}}
\put(-78.,60.){\makebox(0,0){\large cusp}}
\put(-57.,101.){\makebox(0,0){\large DIRAC}}
\put(-69.,104.){\vector(1,0){48.0}}
\put(-69.,104.){\vector(-1,0){29.0}}
\end{picture}
}
\end{center}
\caption{
$68\%$ confidence level ellipses corresponding to the final results of the 
present paper (small solid line ellipse: fit with the ChPT constraint (see Eq.~(\ref{ChPT}));
large solid line  ellipse: fit using $a_0-a_2$ and $a_2$ as independent parameters),
and from $K_{e4}$ decay \cite{Bloch-Devaux:2008conf} 
(small dashed line ellipse: fit with the ChPT constraint; large dashed line ellipse:
fit using $a_0$ and $a_2$ as independent parameters). 
Vertical lines: central value from the DIRAC experiment \cite{Adeva:2005pg} 
(dotted line) and error limits (dashed lines). 
The 1-sigma theoretical band allowed by the ChPT constraint (see Eq.~(\ref{ChPT})) 
is shown by the dotted curves.
}

\label{a2_vs_a02}
\end{figure}

Previous determinations of the $\pi \pi$ scattering lengths have also
relied on the measurement of $K^{\pm} \rightarrow \pi^+\pi^-e^{\pm}\nu_e$ 
($K_{e4}$) decay. Fig. \ref{a2_vs_a02} compares our results 
(Eqs.~(\ref{a0a2_uncon}, \ref{a2_uncon})) with the results
from the most recent analysis of a large sample of $K_{e4}$ decays, also
collected by the NA48/2 collaboration \cite{Bloch-Devaux:2008conf}.

If we use the ChPT constraint (see Eq.~(\ref{ChPT})), we obtain 
(see fit $\fitG$ of Table \ref{femtab1})
\begin{eqnarray}
\nonumber
(a_0-a_2) m_+ = 0.2633\pm0.0024(stat.)\pm \\
0.0014(syst.)\pm0.0019(ext.).
\label {a0a2_link}
\end{eqnarray}
For this fit the theoretical uncertainty affecting the value of $a_0-a_2$ 
is estimated to be $\pm 2\%$ ($\pm 0.0053$) from a recent study of the 
effect of adding three-loop diagrams to the $\kcnn$ decay
amplitude \cite {Gallorini:Thesis:2008} in the frame of
the CI formulation \cite{Cabibbo:2005ez} (the goals of this study
included a more precise estimate of the theoretical uncertainties
affecting the $\pi \pi$ scattering lengths).
This theoretical uncertainty is smaller than that affecting the result
of the fit with $a_0-a_2$ and $a_2$ as free parameters, because
the theoretical uncertainty on $a_2$ becomes negligible when using the
ChPT constraint.

The $68\%$ confidence level ellipse corresponding to the result given
by Eq.~(\ref{a0a2_link}) is also shown in Fig. \ref{a2_vs_a02}, together
with a fit to the $K_{e4}$ data which uses the same ChPT constraint.
The $a_0-a_2$ vs $a_2$ correlation coefficient for this figure has been 
calculated taking into account statistical, systematic and external covariances. 
Its value is  $-0.774$, while the statistical correlation alone is $-0.839$ (see Table \ref{berncorr_E}).

\section*{Summary and conclusions}
\label{conclusions}

We have studied the $\pi^0 \pi^0$ invariant mass distribution measured
from the final sample of $6.031\times 10^7$ $\kcnn$ fully reconstructed decays
collected by the NA48/2 experiment at the CERN SPS. As first observed
in this experiment \cite{Batley:2005ax}, this distribution shows a
cusp-like anomaly at $M_{00}=2m_+$ which is interpreted as an
effect due mainly to the final state charge-exchange scattering process
$\pi^+ \pi^- \rightarrow \pi^0 \pi^0$ in $\kccc$ decay
\cite{Budini:1961,Cabibbo:2004gq}.

Good fits to the $\mm2$ distribution have been obtained using two
different theoretical formulations 
\cite{Cabibbo:2005ez} and \cite{Colangelo:2006va,Bissegger:2008ff}, 
all including next-to-leading order
rescattering terms. We use the results of the fit to the formulation which includes
radiative corrections \cite{Bissegger:2008ff} to determine the difference 
$a_0-a_2$, which enters in the leading-order rescattering term, and
$a_2$, which enters in the higher-order rescattering terms, where $a_0$
and $a_2$ are the $I=0$ and $I=2$ S-wave $\pi \pi$ scattering lengths,
respectively. These values are given in Eqs. (\ref{a0a2_uncon}) and 
(\ref{a2_uncon}), while Eq.~(\ref{a0a2_link}) gives the result from a fit 
that uses the constraint between $a_2$ and $a_0$ predicted by analyticity 
and chiral symmetry \cite{Colangelo:2001sp} (see Eq.~(\ref{ChPT})).

As discussed in Section \ref{final}, our results agree with the values
of the $\pi \pi$ scattering lengths obtained from the study of $K_{e4}$
decay \cite{Bloch-Devaux:2008conf}, which have
errors of comparable magnitude. The value of $a_0-a_2$ as quoted
in Eqs. (\ref{a0a2_uncon}) and (\ref{a0a2_link}) are also in agreement
with theoretical calculation performed in the framework
of Chiral Perturbation Theory \cite{Colangelo:2000jc,Colangelo:2001df},
which predict $(a_0-a_2)m_+ = 0.265 \pm 0.004$.

We finally note a major difference between $\kccc$ and $\kcnn$ decays.
In the case of $\kccc$ decay there is no cusp singularity in the
physical region because the invariant mass of any pion pair is always
 $\ge 2m_+$. As a consequence, rescattering effects can be
reabsorbed in the values of the Dalitz plot parameters 
$g$, $h$, $k$ obtained from fits without rescattering, such as those
discussed in ref. \cite{Batley:2007md}. On the contrary, a correct
description of the $\kcnn$ Dalitz plot is only possible if rescattering
effects are taken into account to the next-to-leading order. Furthermore, 
the values of the parameters $g_0$, $h_0$, $k_0$ which describe the 
weak $\kcnn$ amplitude at tree level depend on the specific theoretical 
formulation of rescattering effects used to fit the data.

\begin{sloppypar}
In a forthcoming paper we propose an empirical
parameterization capable of giving a description of the $\kcnn$ Dalitz plot,
which does not rely on any $\pi \pi$ rescattering mechanisms,
but nevertheless reproduces the cusp anomaly at $M_{00}=2m_+$. This
parameterization is useful for computer simulations of $\kcnn$ decay
requiring a precise description of all Dalitz plot details.
\end{sloppypar}

\section*{Acknowledgements}
We gratefully acknowledge the CERN SPS accelerator and beam-line staff
for the excellent performance of the beam. We thank the technical staff of
the participating laboratories and universities for their effort in the
maintenance and operation of the detectors, and in data processing.
We are grateful to G. Isidori for valuable discussions on the fitting
procedure. It is also a pleasure to thank G. Colangelo, J. Gasser, B. Kubis and
A. Rusetsky for illuminating discussions and for providing the
computer code to calculate the $\kccc$ and $\kcnn$ decay amplitudes in the
framework of the Bern-Bonn formulation.

\section*{Appendix: Measurement of the \boldmath $k_0$ \unboldmath parameter}

In order to measure the $k_0$ parameter which describes the $v^2$ dependence
of the weak amplitude for $\kcnn$ decay at tree level (see Eq.(\ref{amp0})),
we have performed fits to the $\pi^{\pm} \pi^0 \pi^0$ Dalitz plot.
Because of technical complications associated with two-dimensional fits,
we do not use the results of these fits to determine the scattering lengths,
but focus mainly on the measurement of $k_0$.

We use two independent methods. In the first method, the Dalitz plot 
is described by two independent variables: $\mm2$ and $\cos(\theta)$, 
where $\theta$ is the angle between the momentum vectors of the 
$\pi^{\pm}$ and one of the two $\pi^0$ in the rest frame of the $\pi^0$ pair 
(with this choice of variables the Dalitz plot has a rectangular physical boundary). 
The $\mm2$ fit interval is identical to the one used for the one-dimensional 
fits described in Sections \ref{ci}, \ref{bb}, but the bin width is increased 
from 0.00015 to 0.0003~\gevmsq, and four consecutive bins around 
$\mm2=4m_+^2$ are excluded. The $\cos(\theta)$ variable is divided into 21 
equal bins from $-1.05$ to $1.05$, but only the interval $-0.85<\cos(\theta)<0.85$ 
(17 bins) is used in the fits.

In order to take into account the distortions of the theoretical Dalitz plot  
due to acceptance and resolution effects, a four-dimensional matrix 
(with dimensions $210 \times 21 \times 210 \times 21$) is
obtained from the Monte Carlo simulation described in Section \ref{mcsimul}.
This matrix is used to transform the true simulated Dalitz plot into an expected
one which can be directly compared with the measured Dalitz plot at
each step of the $\chi^2$ minimization.

\begin{figure}
\begin{center}
\resizebox{1.0\columnwidth}{!}{%
\setlength{\unitlength}{1mm}
\begin{picture}(85.,85.)
\includegraphics[width=85mm]{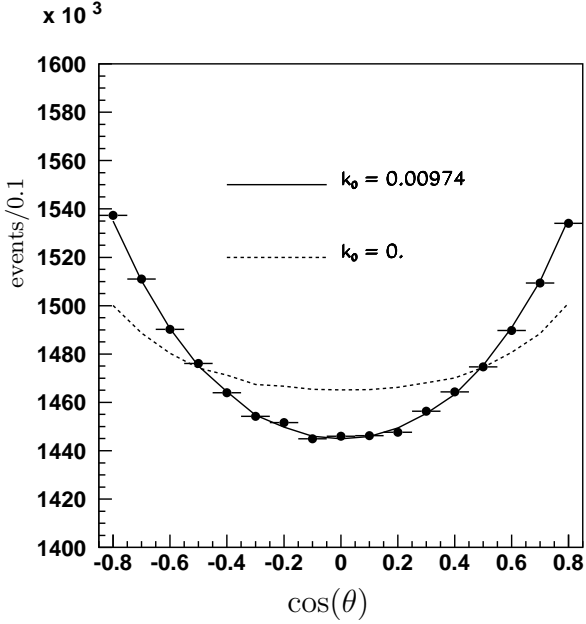}
\put(-42.,1.){\makebox(0,0){\large $\cos(\theta)$}}
\put(-83.,50.){\makebox(0,0){\rotatebox{90}{events/0.1}}}
\end{picture}
}
\end{center}
\caption{Projections of the $\kcnn$ Dalitz plot onto the $\cos(\theta)$ axis
(see text). Full circles: data. Dashed (full) line: best fit to the
CI formulation \cite{Cabibbo:2005ez} with $k_0=0$ ($k_0=0.00974$).}
\label{kvterm}
\end{figure}

Fits to the CI formulation \cite{Cabibbo:2005ez} are
performed with a fixed value $a_2=-0.044$. If the $k_0$ parameter is
kept fixed at zero, the fit quality is very poor ($\chi^2 = 4784.4$ for 1237 degrees 
of freedom); however, if $k_0$ is used as a free parameter in the fit,
the best fit value is $k_0 = 0.00974 \pm 0.00016$, and $\chi^2 = 1223.5$
for 1236 degrees of freedom. The results of these two fits are shown
in Fig. \ref{kvterm}, where the data and best fit Dalitz plots are projected 
onto the $\cos(\theta)$ axis.

A simultaneous fit to the Dalitz plot from $\kcnn$ decay and to the  
$M_{\pm \pm}^2$ distribution from $\kccc$ decay is performed in the frame of 
the BB formulation \cite{Colangelo:2006va} using the constraint between 
$a_2$ and $a_0$ predicted by analyticity and chiral symmetry 
(see Eq.(\ref{ChPT})). The best fit gives $k_0 = 0.00850 \pm 0.00014$, 
with $\chi^2 = 1975.5$ for 1901 degrees of freedom. The difference between
the $k_0$ value so obtained and that obtained from a fit to the
CI formulation \cite{Cabibbo:2005ez} is due to the rescattering 
contributions which are different in the two formulations. When radiative
corrections are included in the fit \cite{Bissegger:2008ff}, 
$k_0$ is practically unchanged (its best fit value is 0.008495), demonstrating
that electromagnetic corrections have a negligible effect on its determination.

The second fitting method is based on the event weighting
technique. In order to study the size of the trigger effect on the fit
parameters, we use a fraction of the data taken with uniform trigger 
conditions and associated with a large minimum bias event sample which
allows a precise evaluation of the trigger efficiency.

The Dalitz plot is described by the $u$ and $|v|$ variables 
(see Eq.(\ref{amp0})), and the intervals $-1.45<u<1.35$ and $|v|<2.8$  
are each sudivided into 50 equal size bins. The fits are performed using
the CI formulation \cite{Cabibbo:2005ez} over a wide region
which excludes only the tails of the distribution ($0 < |v| < 0.9\ v_{max}$, 
$u < 0.9$). All bins around the cusp point are included, and pionium
formation is taken into account by multiplying the theoretical $\kcnn$
decay probability by the factor 1.055 in the interval
$|\mm2-4m_+^2| < 0.000075$~\gevmsq. The fits are performed with 
a fixed value $a_2=-0.044$.

In the fits we use the Dalitz plots distributions of the selected 
events, corrected (or not corrected) for the trigger efficiency, and
of a corresponding subsample of $\sim2.8 \times 10^7$ simulated events 
generated with a simple matrix element
${\cal M}_{sim}$ without rescattering effects and with fixed
values of $g_0$, $h_0$ and $k_0$. At every iteration in the $\chi^2$ 
minimization, each simulated event is reweighted by the ratio 
$\frac{|{\cal M}|^2}{|{\cal M}_{sim}|^2}$, where ${\cal M}$ is the
matrix element which includes rescattering and is calculated with the new 
fitting parameters, and both ${\cal M}$ and ${\cal M}_{sim}$ are calculated 
at the generated $u$, $|v|$ values. The simulated events so weighted are
then rebinned, and their two-dimensional $u,|v|$ distribution is compared
with that of the data. 

A good fit ($\chi^2 = 1166$ for 1257 degrees of freedom) is obtained when
the trigger efficiency is taken into account, giving
$k_0 = 0.00966 \pm 0.00018$. If the trigger effect is ignored,
the $\chi^2$ value is somewhat worse ($\chi^2 = 1276$)
and we obtain $k_0 = 0.01010 \pm 0.00017$. This result demonstrates
that the trigger effect is important for the wide region of the Dalitz
plot used in the fit, increasing the measured $k_0$ by $\approx 0.0004$.

The data used in these fits overlap only partially with the data used
in the fit to the CI formulation \cite{Cabibbo:2005ez} performed 
using the first method and discussed above, but the results have almost
equal statistical errors. We average the two results from the fits
without trigger correction, obtaining $k_0 = (0.00974 + 0.01010)/2 = 0.0099$.
We take the statistical error of one of them as the statistical error of the 
measured $k_0$ value, and conservatively take one half of the difference 
between them as the contribution to the systematic error due to the
different fitting techniques. As mentioned above, the trigger correction 
shifts the $k_0$ central value by $-0.0004$. Because this effect is measured
only with a partial data sample, we also add it in quadrature to the 
systematic error. So our measurement of $k_0$ in the frame of the
CI rescattering formulation \cite{Cabibbo:2005ez} gives

\begin{eqnarray}
\nonumber
k_0 = 0.0095 \pm 0.00017(stat.) \pm 0.00048(syst.) \\
\nonumber
= 0.0095 \pm 0.0005.
\end{eqnarray}

For most of the one-dimensional fits discussed in the present paper we
do not apply any trigger correction, so here we use the effective value 
$k_0 = 0.0099$ for the fits to the CI formulation
\cite{Cabibbo:2005ez}, and $k_0=0.0085$ for the fits to the BB
formulation \cite{Colangelo:2006va,Bissegger:2008ff}. Since $k_0$
is kept fixed in those fits, we check the variations of all the best fit
parameters by varying $k_0$ within the limits defined by its full
error. These variations are listed in Tables \statsys, where they are
denoted as ``$k_0$ error''.

\addcontentsline{toc}{section}{References}
\bibliographystyle{epj}
\bibliography{NA48CUSP}


\end{document}